\documentclass[a4paper,12pt,twoside]{article} 
\usepackage{a4wide} 
\usepackage{amsmath}
\usepackage{amsfonts}
\usepackage{amssymb}
\usepackage{graphicx}
\usepackage{subfigure}
\usepackage{float}
\usepackage{algorithm}
\usepackage{algorithmic}
\usepackage{appendix}
\usepackage{epsfig,subfigure} 
\usepackage{natbib} 
\usepackage{color} 
\usepackage{bm}
\usepackage{hyphenat} 
\usepackage{caption}
\usepackage{lineno}
\usepackage{bm}

\definecolor{myBlue}{rgb}{0,0,0.56} \definecolor{myRed}{rgb}{0.5,0,0}

\makeatletter

\newcommand{\Rmnum}[1]{\expandafter\@slowromancap\romannumeral #1@}
\makeatother

\begin{document}

\title{Toward an enhanced Bayesian estimation framework for multiphase flow soft-sensing}

\author{Xiaodong Luo\thanks{\textit{Corresponding author address:}
				International Research Institute of Stavanger (IRIS), Thorm{\o}hlens Gate 55, 5008 Bergen, Norway.
				{E-mail: xiaodong.luo@iris.no}}, Rolf J. Lorentzen, Andreas S. Stordal and Geir N{\ae}vdal \\[+0.5cm]
International Research Institute of Stanvanger (IRIS), 5008 Bergen, Norway }
\date{}
\maketitle

\begin{abstract}
In this work the authors study the multiphase flow soft-sensing problem based on a previously established framework. There are three functional modules in this framework, namely, a transient well flow model that describes the response of certain physical variables in a well, for instance, temperature, velocity and pressure, to the flow rates entering and leaving the well zones; a Markov jump process that is designed to capture the potential abrupt changes in the flow rates; and an estimation method that is adopted to estimate the underlying flow rates based on the measurements from the physical sensors installed in the well.

In the previous studies, the variances of the flow rates in the Markov jump process are chosen manually. To fill this gap, in the current work two automatic approaches are proposed in order to optimize the variance estimation. Through a numerical example, we show that, when the estimation framework is used in conjunction with these two proposed variance-estimation approaches, it can achieve reasonable performance in terms of matching both the measurements of the physical sensors and the true underlying flow rates. 

\end{abstract}
\section{Introduction}
\label{sec:introduction} 

\noindent 
Smart wells are advanced operation facilities used in modern oil and gas fields. Typically, a smart well is equipped with downhole sensors and inflow control valves (ICVs). The ICVs are used to control the inflow rates of the fluids in the well, which thus requires the information of the inflow rates. Normally, direct flow rate sensors are not often deployed in a smart well, since they are expensive to use, yet the resulting measurements may be unreliable due to the high risk of sensor malfunction or failure. Therefore, in practice it is more customary to install downhole sensors that collect and transmit other types of data, e.g., pressure and temperature. 
Based on the collected data, one can estimate the flow rates in the well. Such an exercise is often called ``soft sensing'' or ``soft metering'' (see, for examples, \citealp{Bloemen2006SPEJsoft,deKruif2008soft,Leskens2008downhole,Lorentzen10Soft,Wrobel2009soft}). Information obtained from soft sensing can then be used to support decision-making, e.g., choosing ICV operation strategies for the purpose of production optimization. In this sense, soft sensing is essential for efficient and profitable production operations.   

Various methods have been proposed in the literature in order to tackle the soft sensing problem based on the measurements collected from downhole sensors. For instance, \citet{Bloemen2006SPEJsoft} used the conventional extended Kalman filter (EKF) for soft sensing in gas-lift wells; \citet{Gryzlov10Estimation,Leskens2008downhole,Lorentzen10Rate,Lorentzen10Soft} adopted the more recently emerged ensemble Kalman filter (EnKF, see, for example, \citealp{Aanonsen-ensemble-2009,Evensen06Data,naevdal2005reservoir}) to estimate the flow rates based on the temperature and/or pressure measurements. To better address the issue of nonlinearity in the process of soft sensing, a more sophisticated method, called the auxiliary particle filter (APF, see, for example, \citealp{Pitt_ASIR}), is employed in a recent work of \citet{LorentzenSPEEstimation}. Apart from the aforementioned (approximate) Bayesian filters, the soft sensing problem can also be solved by alternative methods, for example, through a certain iterative algorithm in the inverse problem theory (see, for example, \citealp{Yoshioka2009new}).

The current work adopts the same framework as that developed in \citet{LorentzenSPEEstimation}. There are three functional modules in this framework, including: (1) a transient multiphase well flow model, in contrast to the conventional steady state models used in many other works (see, for example, \citealp{Azim12Novel}). Here, the well flow model simulates the behaviour of a smart well with given flow rates and other well conditions; (2) a Markov jump process that describes how the flow rates change with time. One reason for \citet{LorentzenSPEEstimation} to adopt this dynamical model with jumps is for its ability to capture the rapid changes in the flow rates, which might take place due to the (sudden) changes of well conditions, for example, in case of water or gas breakthrough, or operational changes of other wells in the field, and so on \citep{LorentzenSPEEstimation}; (3) an estimation method based on the APF that estimates the underlying flow rates, based on available measurements from downhole sensors. Note that in general one can replace the APF by any other suitable estimation method, e.g., those mentioned in the preceding paragraph.  

By using the aforementioned framework, satisfactory soft sensing results are obtained in \citet{LorentzenSPEEstimation}. One of the remaining issues in \citet{LorentzenSPEEstimation} is that, when using the Markov jump process, the variances of the process noise are manually chosen, which are thus not guaranteed to be optimal in general situations. In the current work, we aim to fill this gap by proposing two methods that are able to calculate the optimal variances based on certain optimality criteria. For more information about the transient well flow model and the APF, readers are referred to \citet{LorentzenSPEEstimation,Lorentzen10Rate,Lorentzen10Soft} and \citet{Pitt_ASIR}, as well as the appendix in this work.

This work is organized as follows. Section \ref{sec:methodology} outlines the problem that we are interested in, and proposes two methods to enhance the estimation framework established previously. Section \ref{sec:example} uses an example to illustrate the performance of the estimation framework that is equipped with the proposed methods. Section \ref{sec:conclusion} concludes the work. For readers' better comprehension of the framework used in this work, a short introduction to the APF is also provided in Appendix \ref{sec:appendix}.     

\section{Methodology} \label{sec:methodology}
\subsection{Problem statement}
Without loss of generality, suppose that the (abstract) transient well flow model is described by the following mathematical equation
\begin{equation} \label{eq:y_obv}
\mathbf{y}_k = \mathcal{H}_k (\mathbf{Q}_k) + \mathbf{v}_k \, ,
\end{equation}
where $\mathbf{y}_k$ is the nominal measurement containing, for instance, the temperature and pressure at the time instant $k$; $\mathcal{H}_k$ is the well flow model; and $\mathbf{Q}_k = [Q_{k,1},\dotsb,Q_{k,r}]^T$ is the rates of inflow or outflow of fluid phases in $r$ well zones. Since in practice the measurements from the physical sensors normally may contain certain errors, an additional term, $\mathbf{v}_k$, is thus in presence in order to account for this possibility. In this work we assume that $\mathbf{v}_k$ follows the normal distribution $N(\mathbf{v}_k; \mathbf{0},\mathbf{W}_k)$, with zero mean and covariance $\mathbf{W}_k$. For simplicity, here we further assume that $\mathbf{W}_k$ is known to us, otherwise it may also be estimated by the methods presented below. 

The Markov jump process of the flow rate at the $i$-th well zone is given by
\begin{equation} \label{eq:rates_model}
Q_{k+1,i} = \theta_{k+1,i}  Q_{k,i} + u_{k,i} \, ,~ i = 1, 2, \dotsb, r \, ,
\end{equation} 
where $\theta_{k+1,i}$ is a multiplier that follows a certain discrete probability distribution, e.g., $Pr(\theta_{k+1,i} = c_{j,i}) = P_{j,i}$ ($j = 1, 2, \dotsb, N_i$) for $N_i$ predefined constants $c_{j,i}$ (specific to the well zone $i$), and $u_{k,i}$ the process noise following the normal distribution $N(u_{k,i}; 0,\sigma_{k,i})$, with the variance $\sigma_{k,i}$ to be estimated. 

A few remarks are in order before we proceed further. Firstly, in the context of data assimilation nomenclature (see, for example, \citealp{Evensen06Data}), Eq. (\ref{eq:rates_model}) represents the dynamical model whose states (the flow rates) are to be estimated, while Eq. (\ref{eq:y_obv}) is the associated observation operator that describes the relation between the model states and the observations. Based on this point of view, the noise term $u_{k,i}$ in Eq. (\ref{eq:rates_model}) can be considered as one of the sources of model errors. In general, the distribution of $u_{k,i}$ is unknown. Here we assume that $u_{k,i}$ follows a Gaussian distribution with zero mean but unknown variance $\sigma_{k,i}$. Later we will discuss how to optimize the choice of $\sigma_{k,i}$ (under a certain optimality criterion) in order to improve the performance of the Bayesian data assimilation algorithm in use (see, for example, Appendix \ref{sec:appendix}). Secondly, with the discrete random variable $\theta_{k+1,i}$ and the Gaussian noise $u_{k,i}$ in Eq. (\ref{eq:rates_model}), it can be verified that the corresponding flow rates $Q_{k+1,i}$ is described by a Gaussian mixture model (GMM). In data assimilation practices, using a GMM to approximate the probability density function (pdf) of a random variable has certain advantages. For instance, from a theoretical point of view, the GMM is one type of the kernels in kernel density estimation (KDE) theory and can be used to approximate a generic non-Gaussian pdf \citep{Silverman1986}\footnote{In accordance with this viewpoint, one may choose other types of kernels in approximation, e.g., by letting $u_{k,i}$ in Eq. (\ref{eq:rates_model}) follow some alternative distributions. An investigation in this aspect will, however, not be carried out in the current work.}. In addition, using a Gaussian noise $u_{k,i}$ in Eq. (\ref{eq:rates_model}) is also convenient for our theoretical development later. For instance, this leads to a convex optimization problem in Section \ref{subsubsec:max_step} that can be more conveniently solved, in contrast to potentially non-convex forms with other distribution choices for $u_{k,i}$.           

In \citet{LorentzenSPEEstimation} the authors adopted an alternative nonlinear and non-Gaussian data assimilation algorithm, namely, the auxiliary particle filter (APF, see \citealp{Pitt_ASIR}), to estimate the flow rates based on available measurements. The difference between the APF and the conventional sequential importance re-sampling particle filter (SIR-PF, see, for example, \citealp{Gordon1993}) is mainly at the re-sampling step. In the SIR-PF, the re-sampling is conducted based on the most recent posterior weights of the particles, while in the APF, measurements ahead (in time) of the particles to be re-sampled are also taken into account, such that after re-sampling the generated particles may tend to match the ``future'' measurements better. For brevity, we outline below the main procedures at the re-sampling step of the APF. Readers are referred to the appendix in this work and \citet{Pitt_ASIR} for more details. 

In the context of flow rate estimation, the re-sampling step in the APF is as follows. Suppose that $\mathbf{Q}_k^j = (Q_{k,1}^j,Q_{k,2}^j,\dotsb,Q_{k,r}^j)^T$ is the $j$-th sample in the set $\{\mathbf{Q}_k^j\}_{j=1}^n$, and $w_k^j$ is the posterior weight associated with $\mathbf{Q}_k^j$. For each $\mathbf{Q}_k^j$, a sample $\boldsymbol{\theta}_{k+1}^j \equiv (\theta_{k+1,1}^j,\theta_{k+1,2}^j,\dotsb,\theta_{k+1,r}^j)^T$ is also drawn based on the distributions of $Pr(\theta_{k+1,i})$ ($i=1,2,\dotsb,r$). Let $\boldsymbol{\Omega}_{k+1}^j = (\theta_{k+1,1}^j \, Q_{k,1}^j, \theta_{k+1,2}^j \, Q_{k,2}^j,\dotsb, \theta_{k+1,n}^j \, Q_{k,n}^j)^T$, then one can compute an intermediate weight $\mu_{k+1}^j \propto w_k^j \, N(\mathbf{y}_{k+1};\mathcal{H}_{k+1} (\boldsymbol{\Omega}_{k+1}^j),\mathbf{W}_{k+1})$, where $N(\mathbf{y}_{k+1};\mathcal{H}_{k+1} (\boldsymbol{\Omega}_{k+1}^j),\mathbf{W}_{k+1})$ denotes the pdf of the Gaussian distribution of $\mathbf{y}_{k+1}$ with mean $\mathcal{H}_{k+1} (\boldsymbol{\Omega}_{k+1}^j)$ and covariance $\mathbf{W}_{k+1}$. Based on $\mu_{k+1}^j$, one applies SIR to the indices $j=1,2,...n$ and obtains a new set of indices $j^s, s = 1,2,\dotsb,n$ to pick up samples from the set $\{\boldsymbol{\Omega}_{k+1}^j\}_{j=1}^n$. The indexed samples $\{\boldsymbol{\Omega}_{k+1}^{j^s}\}_{j=1}^n$ has equal weights and is used to generate new flow rates samples $Q_{k+1,i}^s$ through $Q_{k+1,i}^s = \Omega_{k+1,i}^{j^s} + u_{k+1,i}$ ($i = 1,2,\dotsb,r$ and $s= 1,2,\dotsb,n$). Finally, the weights $w_{k+1}^s$ associated with the sample $\mathbf{Q}_{k+1}^s = (Q_{k+1,1}^s,Q_{k+1,2}^s,\dotsb,Q_{k+1,r}^s)^T$ is computed by $w_{k+1}^s \propto N(\mathbf{y}_{k+1};\mathcal{H}_{k+1} (\mathbf{Q}_{k+1}^s),\mathbf{W}_{k+1}) / N(\mathbf{y}_{k+1};\mathcal{H}_{k+1} (\boldsymbol{\Omega}_{k+1}^{j^s}),\mathbf{W}_{k+1})$.                
   
In the course of generating samples $\mathbf{Q}_{k+1}^s$ through the formula $Q_{k+1,i}^s = \Omega_{k+1,i}^{j^s} + u_{k+1,i}$, the variances $\sigma_{k+1,i}$ of the Gaussian random variables $u_{k+1,i}$ are chosen manually in \citet{LorentzenSPEEstimation}, which implies that in general there needs to be a trial-and-error process, and that the optimality of the final chosen values may not be clear. 
In this work we consider two automatic approaches that can be used to optimize, in the sense to be specified later, the choices of $\sigma_{k+1,i}$ based on available measurements. In the first approach, the optimal variances are considered constant in time, and their estimation is obtained by minimizing a cost function with respect to the collected measurements in a given time interval. This type of estimation approach can be classified as a ``fixed interval smoother'' \citep{Simon2006}, therefore for reference it is called the ``fixed interval smoothing approach'' in the current work. In the second approach, the optimal variances are updated with time, while the estimation is carried out by minimizing a cost function with respect to the measurements up to (and including) certain fixed time steps ahead of the variances to be estimated. This corresponds to a ``fixed lag smoother'' \citep{Simon2006}, hence is called the ``fixed lag smoothing approach''.    

\subsection{The fixed interval smoothing approach}
The fixed interval smoothing approach is based on the idea in \citet{LorentzenSPEEstimation}. For notational convenience, let $\mathbf{Y}^o_{0:K} \equiv \{\mathbf{y}^o_0,\mathbf{y}^o_1,\dotsb,\mathbf{y}^o_K\}$ denote the collection of a set of observed measurements $\mathbf{y}^o_{\bullet}$ from time instant 0 to $K$, with $K$ being fixed, and $\bm{\sigma} \equiv (\sigma_1,\dotsb,\sigma_r)^T$ a vector containing the variances of the Markov jump process in different well zones. Our objective is to estimate a constant variance (vector) $\bm{\sigma}$ that minimizes a certain cost function, as will be explained below.

To construct the cost function, we first denote  by $p(\mathbf{Y}^o_{0:K}|\bm{\sigma})$ the joint probability density function (pdf) of $\mathbf{Y}^o_{0:K}$ conditioned on $\bm{\sigma}$. Then the optimal $\bm{\sigma}^{opt}$ is taken as the one that solves the following minimization problem, i.e.,
\begin{linenomath*}
\begin{equation} \label{eq:min_FI}
\begin{split}
& \bm{\sigma}^{opt} = \underset{\bm{\sigma} \geq 0}{\min} \, C(\bm{\sigma}) \, , \\
& C(\bm{\sigma}) \equiv - \log p(\mathbf{Y}^o_{0:K}|\bm{\sigma}) = - \sum\limits_{k = 1}^{K} \log p(\mathbf{y}^o_{k}|\mathbf{Y}^o_{0:k-1};\bm{\sigma}) \, , 
\end{split}
\end{equation}
\end{linenomath*}
where in the second line of Eq. (\ref{eq:min_FI}) the identity $p(\mathbf{Y}^o_{0:K}|\bm{\sigma}) = \prod\limits_{k=1}^{K} p(\mathbf{y}^o_{k}|\mathbf{Y}^o_{0:k-1};\bm{\sigma})$ is used. 

Note that, in Eq. (\ref{eq:min_FI}), $C(\bm{\sigma})$ is the sum of the terms $-\log p(\mathbf{y}^o_{k}|\mathbf{Y}^o_{0:k-1};\bm{\sigma})$ at individual time instants, in which we have assumed that $\bm{\sigma}$ is constant over time\footnote{In principle, one may choose to let $\bm{\sigma}$ change with time (e.g., by replacing $p(\mathbf{y}^o_{k}|\mathbf{Y}^o_{0:k-1};\bm{\sigma})$ with $p(\mathbf{y}^o_{k}|\mathbf{Y}^o_{0:k-1};\bm{\sigma}_k)$) in Eq. (\ref{eq:min_FI}). In doing so, however, the search space of the minimization problem is larger and the optimization process becomes more expensive.}. Therefore, in order to optimize $\bm{\sigma}$, the observations in a fixed time interval (e.g., from 0 to $K$ in Eq. (\ref{eq:min_FI})) need to be used. In the literature, the corresponding data assimilation scheme is often called fixed interval smoother (see, for example, \citealp{Simon2006}), in which one needs to wait until all the observations in the specified time interval arrive before starting the estimation. Because of the relatively long waiting time (e.g., 100 minutes for our case study in Section \ref{sec:example}), the fixed interval smoothing approach is often used for offline analysis.   

Further note that
\begin{linenomath*}
\begin{equation} \label{eq:cond_pdf_FI}
p(\mathbf{y}^o_{k}|\mathbf{Y}^o_{0:k-1};\bm{\sigma}) = \int p(\mathbf{y}^o_{k}|\mathbf{Q}_k) p(\mathbf{Q}_k|\mathbf{Y}^o_{0:k-1};\bm{\sigma}) d\mathbf{Q}_k \, , 
\end{equation}
\end{linenomath*}       
where $p(\mathbf{y}^o_{k}|\mathbf{Q}_k)$ is the conditional pdf of $\mathbf{y}^o_{k}$ on a given flow rates vector $\mathbf{Q}_k$, and can be determined through Eq. (\ref{eq:y_obv}); $p(\mathbf{Q}_k|\mathbf{Y}^o_{0:k-1};\bm{\sigma})$ is the prior (or predictive) pdf of the flow rates $\mathbf{Q}_k$, conditioned on $\bm{\sigma}$ and the past measurements up to (and including) time instant $k-1$. For a given $\bm{\sigma}$, $p(\mathbf{Q}_k|\mathbf{Y}^o_{0:k-1};\bm{\sigma})$ can be obtained from a Bayesian filter. For instance, in the context of particle filtering, 
\begin{linenomath*}
\begin{equation} \label{eq:prior_pdf_FI}
p(\mathbf{Q}_k|\mathbf{Y}^o_{0:k-1};\bm{\sigma}) \approx \sum\limits_{h=1}^{n} \, w_h^{pr} \, \delta(\mathbf{Q}_k - \mathbf{Q}_{k,h}) \, ,
\end{equation}
\end{linenomath*}   
where $\{\mathbf{Q}_{k,h}\}_{h=1}^n$ is a set of $n$ samples (often called ``particles'') of the flow rates $\mathbf{Q}_k$; $\{w_h^{pr}\}_{h=1}^n$ is a set of \textit{a priori} weights associated with the particles, before the measurement $\mathbf{y}^o_{k}$ is incorporated in estimation; and $\delta(\bullet)$ represents the Dirac delta function, whose value is $+\infty$ at the origin, and 0 elsewhere.      

Substituting Eqs. (\ref{eq:cond_pdf_FI}) and (\ref{eq:prior_pdf_FI}) into Eq. (\ref{eq:min_FI}), one obtains an approximate cost function, in terms of
\begin{linenomath*}
\begin{equation} \label{eq:approx_cost_FI}
\tilde{C}(\bm{\sigma}) \approx - \sum\limits_{k = 1}^{K} \log \left( \sum\limits_{h=1}^{n} \, w_h^{pr} \, p(\mathbf{y}^o_{k}|\mathbf{Q}_{k,h}) \right) \, . 
\end{equation}
\end{linenomath*}
The impact of $\bm{\sigma}$ on the cost function is implicitly made through Eq. (\ref{eq:rates_model}), which influences the generated particles $\mathbf{Q}_{k,h}$. With the approximate cost function, a suitable optimization algorithm can then be employed to optimize the value of $\bm{\sigma}$.  

\subsection{The fixed lag smoothing approach}      
In the fixed lag smoothing approach, the variance $\bm{\sigma}_k$ changes with time, instead of being constant. The amount of measurements used to estimate $\bm{\sigma}_k$ is , in most cases, also different from that in the fixed interval smoothing approach. In general, if one wants to estimate $\bm{\sigma}_k$, it is the collection of the measurements $\mathbf{Y}^o_{0:k+l} \equiv \{\mathbf{y}^o_0,\mathbf{y}^o_1,\dotsb,\mathbf{y}^o_{k+l}\}$  ($k+l \leq K$) that is used, with $l$ being a fixed (often small) positive integer (but note that the sum $k+l$ changes with $k$). Compared with the fixed interval smoothing approach, its fixed lag counterpart tends to be more suitable for real time, online estimation tasks, since it only needs to wait for the $l$-step-ahead observations (i.e., observations from time instant $k+1$ to $k+l$) to arrive before the estimation of $\bm{\sigma}_k$ starts, rather than all the observations from time instant 0 to $K$ as in the fixed interval smoothing approach. In the current study, $l=1$ is chosen in the analysis below, so that the method is also called lag-1 smoothing approach hereafter\footnote{Under this choice, the waiting time for the one-step-ahead observation is 2 minutes for our case study in Section \ref{sec:example}, which appears more practical, in contrast to 100 minutes in the fixed interval smoothing approach.}. In general, one may also consider longer lags, which, however, is beyond the scope of the current work.

\begin{figure*}
\centering
\includegraphics[scale=0.4]{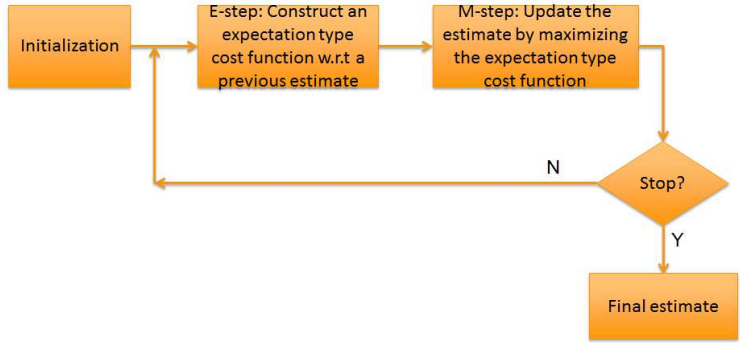}
\caption{\label{fig:EM_framework} An outline of the workflow in the EM algorithm.}
\end{figure*}  

The optimization of $\bm{\sigma}_k$ is done in a way similar to the expectation-maximization (EM) algorithm \citep{Dempster1977maximum}. The EM algorithm is an iterative method and consists of two steps at each iteration. At the expectation step (E-step), one constructs a cost function that can be interpreted as the expectation of a certain quantity. The cost function depends on the parameter(s), say $\bm{\sigma}_k^j$, that will be estimated at the iteration step $j$, and it is also conditioned on the known parameter value, say $\bm{\sigma}_k^{j-1}$, that is obtained at the previous $(j-1)$-th iteration step. Given the cost function, the maximization step (M-step) then takes place in order to estimate an optimal value $\bm{\sigma}_k^j$ that maximizes the cost function value. Fig. \ref{fig:EM_framework} outlines the workflow of the EM algorithm. At the initialization step one specifies the initial parameter value, say $\bm{\sigma}_k^0$. One then constructs an expectation-type cost function that is dependent on $\bm{\sigma}_k^1$ and conditioned on $\bm{\sigma}_k^0$. By maximizing the cost function, one obtains the optimal parameter value of $\bm{\sigma}_k^1$. One then checks if the stopping condition is met. If not, a new iteration starts, in which one constructs a new cost function that is dependent on the parameter $\bm{\sigma}_k^2$ to be estimated, and conditioned on the optimal value of $\bm{\sigma}_k^1$ obtained at the previous step, and so on. 

The stopping condition in our implementation is either (a) the relative Euclidean norm change of two consecutively estimated parameters, in term of $\Vert \bm{\sigma}_k^j - \bm{\sigma}_k^{j-1} \Vert_2 / \Vert \bm{\sigma}_k^{j-1} \Vert_2$, is lower than a pre-chosen threshold; or (b) the maximum iteration number is reached. The estimate of $\bm{\sigma}_k$ is then taken as the final one of the iteration process. In what follows we describe the E-step and M-step with more details. 

\subsubsection{The expectation step} 
Without loss of generality, suppose that the variance $\bm{\sigma}_k$ at time instant $k$ needs to be estimated. 
Also suppose that an estimate $\bm{\sigma}_k^{j-1}$ is already obtained at the $(j-1)$-th iteration of the EM algorithm, so that the E-step aims to construct a cost function that depends on $\bm{\sigma}_k^{j}$. 

In line with the choice $l=1$, the measurements $\mathbf{Y}^o_{0:k+1} \equiv \{\mathbf{y}^o_0,\mathbf{y}^o_1,\dotsb,\mathbf{y}^o_{k+1}\}$ are involved in constructing the cost function, which is given by 
\begin{linenomath*}
\begin{equation} \label{eq:E_cost_FL}
\begin{split}
E(\bm{\sigma}_k^j | \bm{\sigma}_k^{j-1})  \equiv \int \, & \log f\left( \mathbf{y}, \mathbf{Q}_k, \mathbf{Q}_{k+1}, \bm{\theta}_{k+1} |  \bm{\sigma}_k^{j}, \mathbf{Y}^o_{0:k} \right) \, f\left( \mathbf{y}, \mathbf{Q}_k, \mathbf{Q}_{k+1}, \bm{\theta}_{k+1} |  \bm{\sigma}_k^{j-1}, \mathbf{Y}^o_{0:k} \right) \\
& \times \delta(\mathbf{y} - \mathbf{y}^o_{k+1}) \, d\mathbf{y} \, d\mathbf{Q}_k \, d\mathbf{Q}_{k+1} \, d\bm{\theta}_{k+1} \, .
\end{split}
\end{equation}
\end{linenomath*}
In Eq. (\ref{eq:E_cost_FL}), $\mathbf{y}$ is a dummy variable with respect to the measurement $\mathbf{y}^o_{k+1}$. Since $\mathbf{y}^o_{k+1}$ is already obtained, the Dirac delta function $\delta(\mathbf{y} - \mathbf{y}^o_{k+1})$ is thus introduced; $\bm{\theta}_{k+1} \equiv (\theta_{k+1,1},\dotsb, \theta_{k+1,r})^T$ is the augmented vector of the multipliers $\theta_{k+1,1},\dotsb, \theta_{k+1,r}$ in the Markov jump process Eq. (\ref{eq:rates_model}). Eq. (\ref{eq:E_cost_FL}) can be interpreted as the expectation of the log likelihood of $\mathbf{y}^o_{k+1}$, $\mathbf{Q}_k$, $\mathbf{Q}_{k+1}$ and $\bm{\theta}_{k+1}$ (conditioned on $\bm{\sigma}_k^{j}$ and $\mathbf{Y}^o_{0:k}$), with respect to the joint pdf of $\mathbf{y}^o_{k+1}$, $\mathbf{Q}_k$, $\mathbf{Q}_{k+1}$ and $\bm{\theta}_{k+1}$ (conditioned on $\bm{\sigma}_k^{j-1}$ and $\mathbf{Y}^o_{0:k}$). 

Let $\bm{\sigma}_k^{\bullet}$ represent either $\bm{\sigma}_k^{j-1}$ or $\bm{\sigma}_k^{j}$, then one can factorize the joint pdf 
\[f\left( \mathbf{y}^o_{k+1}, \mathbf{Q}_k, \mathbf{Q}_{k+1}, \bm{\theta}_{k+1} |  \bm{\sigma}_k^{\bullet}, \mathbf{Y}^o_{0:k} \right)\] 
as follows:
\begin{linenomath*}
\begin{equation} \label{eq:pdf_factor_FL}
\begin{split}
f\left( \mathbf{y}^o_{k+1}, \mathbf{Q}_k, \mathbf{Q}_{k+1}, \bm{\theta}_{k+1} |  \bm{\sigma}_k^{\bullet}, \mathbf{Y}^o_{0:k} \right) = 
f\left( \mathbf{y}^o_{k+1} | \mathbf{Q}_{k+1} \right) \, 
f\left( \mathbf{Q}_{k+1} | \mathbf{Q}_k, \bm{\theta}_{k+1}, \bm{\sigma}_k^{\bullet} \right) \:
f\left( \mathbf{Q}_k |  \mathbf{Y}^o_{0:k} \right) \,
f\left( \bm{\theta}_{k+1} \right) \, . 
\end{split}
\end{equation}
\end{linenomath*}           
In Eq. (\ref{eq:pdf_factor_FL}), the conditional pdf $f\left( \mathbf{y}^o_{k+1} | \mathbf{Q}_{k+1} \right)$ can be calculated based on the transient well flow model Eq. (\ref{eq:y_obv}); the conditional pdf $f\left( \mathbf{Q}_{k+1} | \mathbf{Q}_k, \bm{\theta}_{k+1}, \bm{\sigma}_k^{\bullet} \right)$ and the pdf (with discrete variables) $f\left( \bm{\theta}_{k+1} \right)$ are determined in the Markov jump process Eq. (\ref{eq:rates_model}); the conditional pdf $f\left( \mathbf{Q}_k |  \mathbf{Y}^o_{0:k} \right)$ is the posterior pdf of the flow rates,  conditioned on the past measurement $\mathbf{Y}^o_{0:k}$ up to (and including) time instant $k$. Similar to the situation in the fixed interval smoothing approach (see Eq. (\ref{eq:prior_pdf_FI})), in the context of particle filtering, $f\left( \mathbf{Q}_k |  \mathbf{Y}^o_{0:k} \right)$ can also be approximated by the empirical posterior pdf of the samples of the flow rates.     

Based on the factorization in Eq. (\ref{eq:pdf_factor_FL}), the log likelihood term in Eq. (\ref{eq:E_cost_FL}) can be re-written as
\begin{linenomath*}
\begin{equation} \label{eq:log_likelihood_FL}
\begin{split}
\log f\left( \mathbf{y}^o_{k+1}, \mathbf{Q}_k, \mathbf{Q}_{k+1}, \bm{\theta}_{k+1} |  \bm{\sigma}_k^{j}, \mathbf{Y}^o_{0:k} \right) = &
\log f\left( \mathbf{Q}_{k+1} | \mathbf{Q}_k, \bm{\theta}_{k+1}, \bm{\sigma}_k^{j} \right) \\
& + \log \left(  f\left( \mathbf{y}^o_{k+1} | \mathbf{Q}_{k+1} \right) \, f\left( \bm{\theta}_{k+1} \right) \, 
f\left( \mathbf{Q}_k |  \mathbf{Y}^o_{0:k} \right)\right) \, . 
\end{split}
\end{equation}
\end{linenomath*}        
The second term on the right hand side of Eq. (\ref{eq:log_likelihood_FL}) is independent of $\bm{\sigma}_k^{j}$. Therefore, at the M-step this term can be ignored for the purpose of optimization.
      
\subsubsection{The maximization step}  \label{subsubsec:max_step} 
Combining Eqs. (\ref{eq:E_cost_FL}) - (\ref{eq:log_likelihood_FL}), at the maximization step one needs to solve the following optimization problem
\begin{linenomath*}
\begin{equation} \label{eq:max_problem}
\begin{split}
\arg\max_{\bm{\sigma}_k^{j}} \tilde{E}(\bm{\sigma}_k^j | \bm{\sigma}_k^{j-1})  \equiv \int \, & \log f\left( \mathbf{Q}_{k+1} | \mathbf{Q}_k, \bm{\theta}_{k+1}, \bm{\sigma}_k^{j} \right) \, 
f\left( \mathbf{y}^o_{k+1} | \mathbf{Q}_{k+1} \right) \, 
f\left( \mathbf{Q}_{k+1} | \mathbf{Q}_k, \bm{\theta}_{k+1}, \bm{\sigma}_k^{j-1} \right) \, \\
& \times 
f\left( \bm{\theta}_{k+1} \right) \, f\left( \mathbf{Q}_k |  \mathbf{Y}^o_{0:k} \right) \, d\mathbf{Q}_k \, d\mathbf{Q}_{k+1} \, d\bm{\theta}_{k+1} \, , 
\end{split}
\end{equation}
\end{linenomath*}   
with the simplified cost function $\tilde{E}(\bm{\sigma}_k^j | \bm{\sigma}_k^{j-1})$ by discarding the irrelevant term in Eq. (\ref{eq:log_likelihood_FL}). Since in general it is difficult to obtain the analytical form of the integral in (\ref{eq:max_problem}), it is customary in practice to approximate that integral through certain Monte Carlo approximations. Specifically, let $\{\mathbf{Q}_{k,h}\}_{h=1}^n$ be a set of flow rates samples at time instant $k$, and $\{w_{h}^{post}\}_{h=1}^n$ the associated posterior weights, then
\begin{linenomath*}
\begin{equation} \label{eq:post_pdf_FL}
f\left( \mathbf{Q}_k |  \mathbf{Y}^o_{0:k} \right) \approx \sum\limits_{h=1}^{n} \, w_h^{post} \, \delta(\mathbf{Q}_k - \mathbf{Q}_{k,h}) \, .
\end{equation}
\end{linenomath*}  
Similarly, let $\{\bm{\theta}_{k,l}\}_{l=1}^m$ be a set of multiplier samples, then 
\begin{linenomath*}
\begin{equation} \label{eq:mutiplier_pdf_FL}
f\left( \bm{\theta}_{k+1} \right) d\bm{\theta}_{k+1} \approx \dfrac{1}{m} \sum\limits_{l=1}^{m} \, \tilde{\delta}(\bm{\theta}_{k+1} - \bm{\theta}_{k+1,l}) \, , 
\end{equation}
\end{linenomath*}  
where $\tilde{\delta}$ is the Kronecker delta function whose value is 1 at the origin and 0 elsewhere. The constant factor $1/m$ is not essential in optimization and is thus dropped later. 

Given a flow rate sample $\mathbf{Q}_{k,h} \equiv (Q_{k,h,1},\dotsb, Q_{k,h,r})^T$ and a multiplier sample $\bm{\theta}_{k+1,l} \equiv (\theta_{k+1,l,1},\dotsb,\theta_{k+1,l,r})^T$, $f\left( \mathbf{Q}_{k+1} | \mathbf{Q}_{k,h}, \bm{\theta}_{k+1,l}, \bm{\sigma}_k^{j-1} \right)$ follows a normal distribution with the mean $ \mathbf{Q}_{k,h} \cdot \bm{\theta}_{k+1,l} \equiv (\theta_{k+1,h,1} Q_{k,l,1}, \dotsb, \theta_{k+1,h,r} Q_{k,l,r})^T$, and the variances being $\sigma_{k,i}^{j-1}$ for $i = 1, \dotsb, r$. For Monte Carlo approximation, note that \[
f\left( \mathbf{Q}_{k+1} | \mathbf{Q}_{k,h}, \bm{\theta}_{k+1,l}, \bm{\sigma}_k^{j-1} \right) \, d\mathbf{Q}_{k+1} =  
\dfrac{f\left( \mathbf{Q}_{k+1} | \mathbf{Q}_{k,h}, \bm{\theta}_{k+1,l}, \bm{\sigma}_k^{j-1} \right)}{N(\mathbf{Q}_{k+1}; \bm{\mu}, \bm{\Sigma})} \: N(\mathbf{Q}_{k+1}; \bm{\mu}, \bm{\Sigma}) \, d\mathbf{Q}_{k+1} \, ,
\]
where $N(\mathbf{Q}_{k+1}; \bm{\mu}, \bm{\Sigma})$ is the multivariate normal distribution with the mean being $\bm{\mu}$ and the covariance being $\bm{\Sigma}$. With these, one draws a set of $s$ samples, say $\{\mathbf{Q}_{k+1,q}\}_{q=1}^s$, from $N(\mathbf{Q}_{k+1}; \bm{\mu}, \bm{\Sigma})$, such that
\begin{linenomath*}
\begin{equation} \label{eq:flow_rates_pdf_FL}
\begin{split}
f\left( \mathbf{Q}_{k+1} | \mathbf{Q}_{k,h}, \bm{\theta}_{k+1,l}, \bm{\sigma}_k^{j-1} \right) & \approx \sum\limits_{q=1}^{s} \,  \delta(\mathbf{Q}_{k+1} - \mathbf{Q}_{k+1,q}) \, \dfrac{f\left( \mathbf{Q}_{k+1} | \mathbf{Q}_{k,h}, \bm{\theta}_{k+1,l}, \bm{\sigma}_k^{j-1} \right)}{N(\mathbf{Q}_{k+1}; \bm{\mu}, \bm{\Sigma})} \, .
\end{split}
\end{equation}
\end{linenomath*}        
It is natural to let $\bm{\mu}$ be the sample mean of $\mathbf{Q}_{k,h} \cdot \bm{\theta}_{k+1,l}$ (with different combinations of $\mathbf{Q}_{k,h}$ and $\bm{\theta}_{k+1,l}$); on the other hand, in our implementation we let the variances of $\bm{\Sigma}$ be 3 times larger than the empirical variances of $\mathbf{Q}_{k,h} \cdot \bm{\theta}_{k+1,l}$, such that it is (more) likely to have certain samples among $\{\mathbf{Q}_{k+1,q}\}_{q=1}^s$ that are relatively far away from the sample mean of $\mathbf{Q}_{k,h} \cdot \bm{\theta}_{k+1,l}$, even in the case that the empirical variances of $\mathbf{Q}_{k,h} \cdot \bm{\theta}_{k+1,l}$ themselves are relatively small. 

Substituting Eqs. (\ref{eq:post_pdf_FL}) - (\ref{eq:flow_rates_pdf_FL}) into (\ref{eq:max_problem}), one obtains a Monte Carlo approximation of $\tilde{E}(\bm{\sigma}_k^j | \bm{\sigma}_k^{j-1})$, in terms of
\begin{linenomath*}
\begin{equation} \label{eq:approx_pdf_FL}
\begin{split}
& \hat{E}(\bm{\sigma}_k^j | \bm{\sigma}_k^{j-1}) = \sum\limits_{h,l,q} \, \Omega_{h,l,q}^j \, \log f\left( \mathbf{Q}_{k+1,q} | \mathbf{Q}_{k,h}, \bm{\theta}_{k+1,l}, \bm{\sigma}_k^{j} \right) \, , \\
& \Omega_{h,l,q}^j = \dfrac{w_h^{post} \, f\left( \mathbf{y}^o_{k+1} | \mathbf{Q}_{k+1,q} \right) \, f\left( \mathbf{Q}_{k+1,q} | \mathbf{Q}_{k,h}, \bm{\theta}_{k+1,l}, \bm{\sigma}_k^{j-1} \right)}{N(\mathbf{Q}_{k+1,q}; \bm{\mu}, \bm{\Sigma})} \, .
\end{split}
\end{equation}
\end{linenomath*}     

In order for $\bm{\sigma}_k^j$ to solve the maximization problem, a necessary condition is that
\begin{linenomath*}
\begin{equation} \label{eq:nec_cond_FL}
\dfrac{\partial \hat{E}(\bm{\sigma}_k^j | \bm{\sigma}_k^{j-1})}{\partial \bm{\sigma}_k^j} = \mathbf{0}. 
\end{equation}
\end{linenomath*}   
Solving (\ref{eq:nec_cond_FL}), one obtains the following iteration formula
\begin{linenomath*}
\begin{equation} \label{eq:iter_formula_FL}
\sigma_{k,i}^{j} = \dfrac{\sum\limits_{h,l,q} \, \Omega_{h,l,q}^j \, (Q_{k+1,q,i} - \theta_{k+1,h,i} Q_{k,l,i})^2}{\sum\limits_{h,l,q} \, \Omega_{h,l,q}^j}, ~\text{for}~i = 1, 2, \dotsb, r \, ,
\end{equation}
\end{linenomath*}    
which is an (iterative) weighted squared error estimation in light of the Markov jump process Eq. (\ref{eq:rates_model}). 

Given Eqs. (\ref{eq:approx_pdf_FL}) and (\ref{eq:iter_formula_FL}), the iteration is done in the following way: the variance estimate $\bm{\sigma}_k^{j-1}$ obtained at the $(j-1)$-th iteration is used to calculate the weights $\Omega_{h,l,q}^j$ according to Eq. (\ref{eq:approx_pdf_FL}); then based on Eq. (\ref{eq:iter_formula_FL}), the variance estimate is updated to $\bm{\sigma}_k^{j} = (\sigma_{k,1}^{j},\dotsb, \sigma_{k,r}^{j})^T $, which can then be used to calculate the weights $\Omega_{h,l,q}^{j+1}$ at the next iteration, and so on. Note that by Eq. (\ref{eq:iter_formula_FL}), the variance estimates are positive definite (almost surely). In addition, when calculating the (iterative) weights in Eq. (\ref{eq:iter_formula_FL}), the component $w_h^{post} \, f\left( \mathbf{y}^o_{k+1} | \mathbf{Q}_{k+1,q} \right)/N(\mathbf{Q}_{k+1,q}; \bm{\mu}, \bm{\Sigma})$ only needs to be calculated once for all, while it is only the component $f\left( \mathbf{Q}_{k+1,q} | \mathbf{Q}_{k,h}, \bm{\theta}_{k+1,l}, \bm{\sigma}_k^{j-1} \right)$ that needs to be iteratively updated. This feature significantly reduces the computational cost of the iteration process.

Finally we note that, since $f\left( \mathbf{Q}_{k+1} | \mathbf{Q}_{k,h}, \bm{\theta}_{k+1,l}, \bm{\sigma}_k^{j} \right)$ follows a normal distribution, it is easy to verify that the Hessian ${\partial^2 \hat{E}(\bm{\sigma}_k^j | \bm{\sigma}_k^{j-1})}/{\partial^2 \bm{\sigma}_k^j}$  of the approximate cost function $\hat{E}(\bm{\sigma}_k^j | \bm{\sigma}_k^{j-1})$ is negative definite, which implies that the iteration formula Eq. (\ref{eq:iter_formula_FL}) would indeed lead to a variance estimation that maximizes the approximate cost function $\hat{E}(\bm{\sigma}_k^j | \bm{\sigma}_k^{j-1})$. 

\section{Example} \label{sec:example}  

In this section we illustrate, through an example, the performance of the APF with the optimized variance estimates obtained from the fix interval/lag smoothing approaches. This example is taken from the case study 2 in \citet{LorentzenSPEEstimation}. Note that, due to some recent technical updates, the transient well flow model we are currently using may behave slightly different than the one previously presented in \citet{LorentzenSPEEstimation}. In what follows we consider two scenarios. In the first one, the well flow model is run with a resolution of 50m (in measured depth, md for short) in order to produce the synthetic observations (pressure and temperature), while in the course of data assimilation later, the well flow model is run with the same resolution (50m). For distinction later, we call this \textit{perfect observation scenario} (POS). In the second scenario, the well flow model is run with a finer resolution of 5m to produce the synthetic observations, while in the course of data assimilation, the well flow model is run with a coarser resolution of 50m. This introduces some additional observation errors to the observation system, apart from the noise term $\mathbf{v}_k$ in Eq. (\ref{eq:y_obv}). For this reason, we call this \textit{imperfect observation scenario} (IOS). In addition, in a few cases later we also consider the possibility that the distribution of the observation error $\mathbf{v}_k$ in data assimilation may mismatch the one used to produce the synthetic observation. Note that our purpose here is to examine the performance of the APF in the IOS, and no bias correction procedure is introduced in data assimilation to account for the imperfection of the observation system. All the experiments below are repeated for a few times (each time with a different seed in the random number generator), and similar results are observed. For conciseness, in what follows we only present the results with respect to one of the random seeds. 

\subsection{The perfect observation scenario (POS)}

\begin{figure*}
\centering
\includegraphics[scale=0.4]{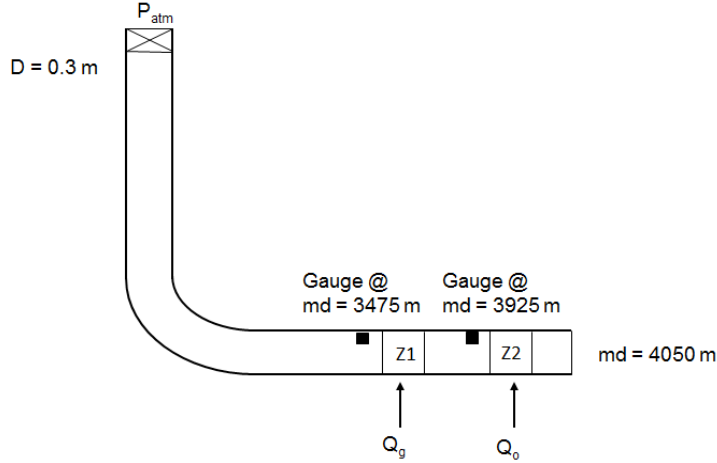}
\caption{\label{fig:well_example} Well configuration.}
\end{figure*}     

In the POS, the geometry of the well is the same as that in \citet{LorentzenSPEEstimation}. As shown in Fig. \ref{fig:well_example}, the well has a vertical part and a horizontal part, connected through an inclined part in-between. The inclined part has a curvature of 0.11 degrees/m, starting from the vertical depth of 1000 m and ending up with the vertical depth of 1500 m. In the horizontal part, there are two influx zones: one locates between md 3500 - 3550 m (labelled as Z1), and the other between 3950 - 4000 m (labelled as Z2). In Z1 there is only gas influx, while in Z2 there is only oil influx. There are two sensors (gauges) installed in the well, with the md being 3475 m and 3925 m, respectively (both gauges are placed 25 m above the influx zone). These sensors collect the temperature and pressure at their locations. In our experimental settings, the reservoir temperature and pressure in Z2 are 335.5 K and $1.5 \times 10^7$ Pa, respectively, while in Z1, they are 325.5 K and $1.4 \times 10^7$ Pa, respectively. For more details, readers are referred to \citet{LorentzenSPEEstimation}.

\begin{figure*}
\centering
\subfigure[True flow rates]{ \label{subfig: flowrates}
\includegraphics[scale=0.4]{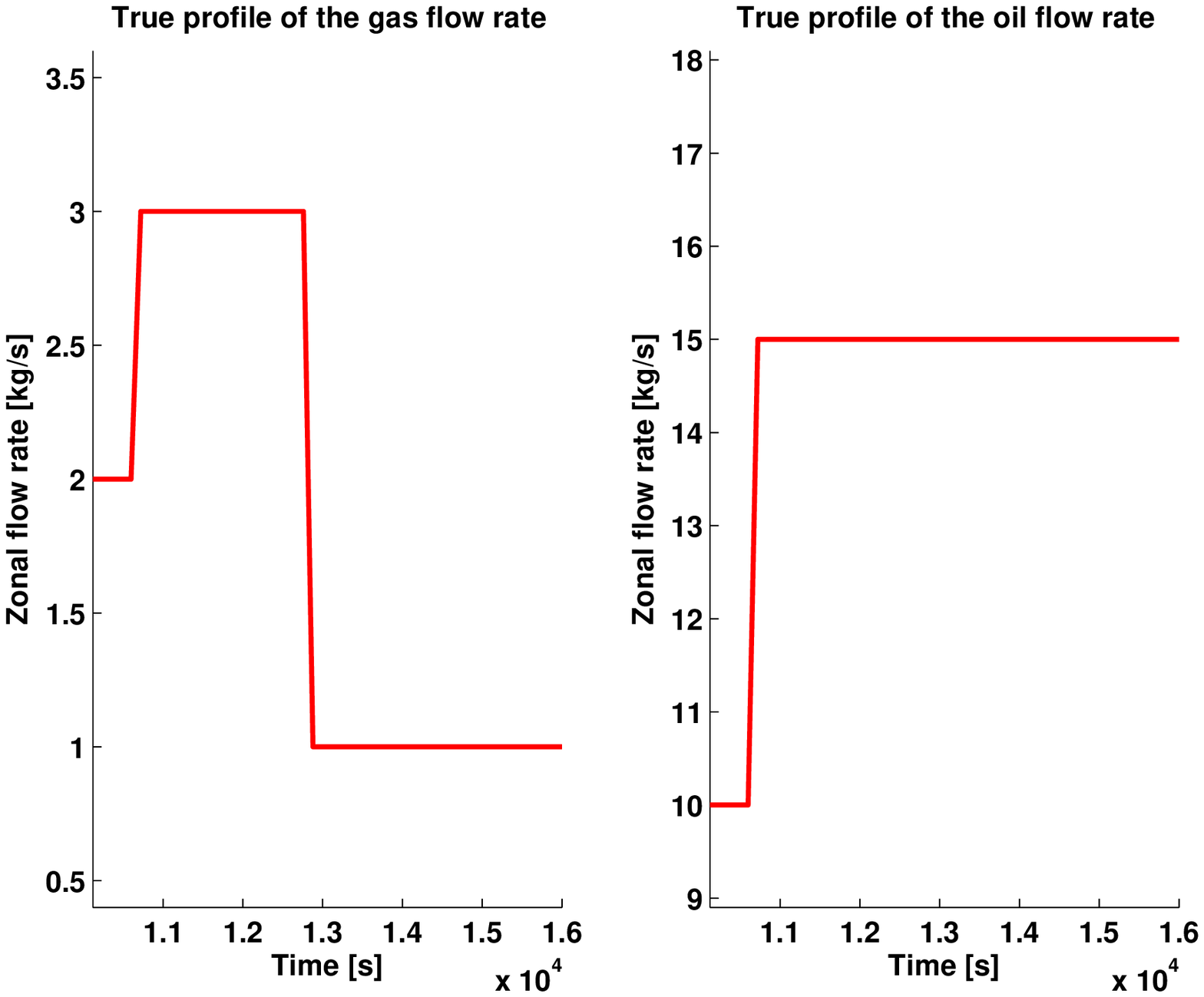}
}
\subfigure[Measured pressure and temperature]{ \label{subfig: temp_press}
\includegraphics[scale=0.4]{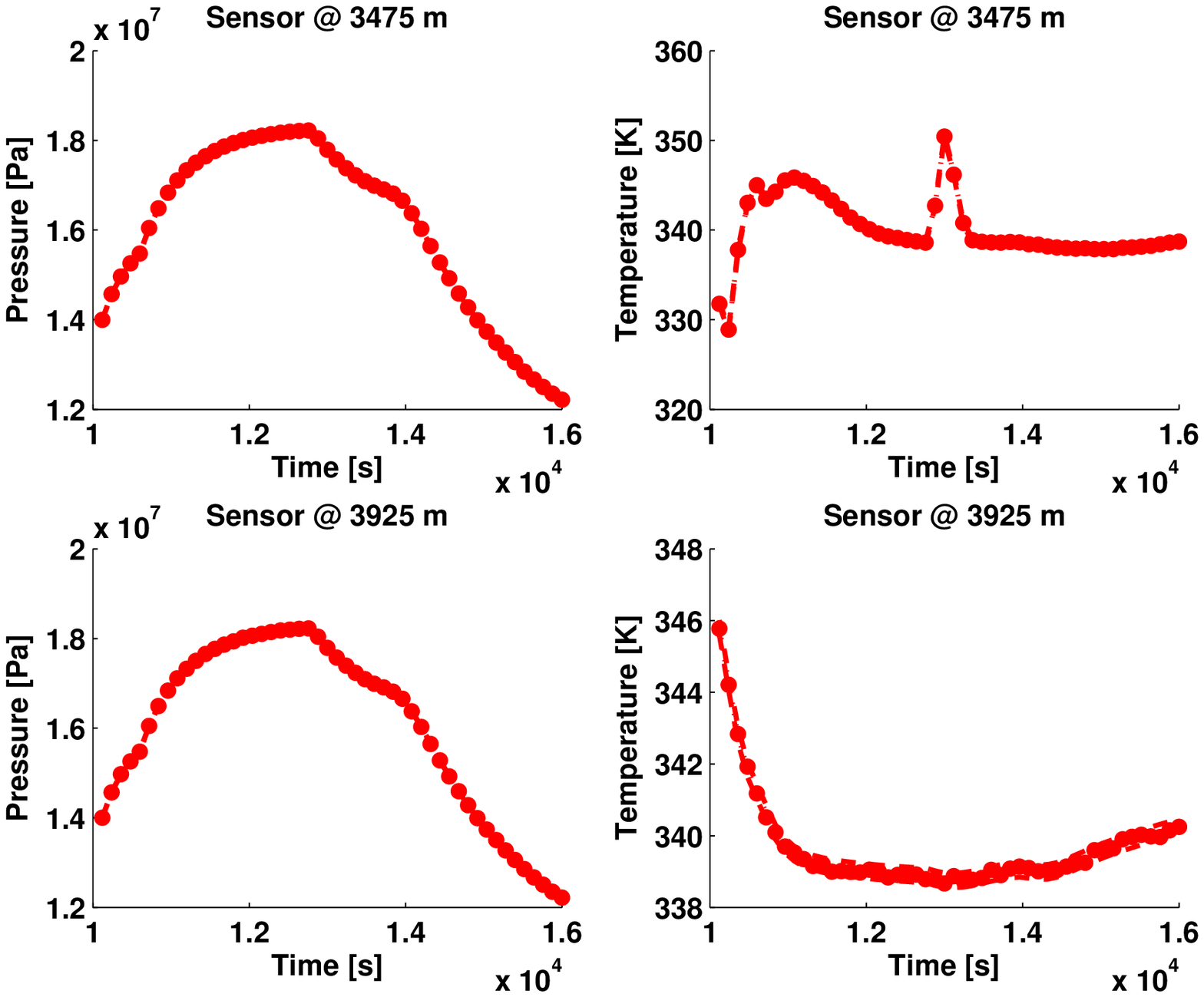}
}
\caption{\label{fig:truth_obs} (a) True gas and oil flow rates in Z1 and Z2 in the POS; (b) Measured pressure and temperature from the gauges in the POS.}
\end{figure*}  
    
Fig. \ref{subfig: flowrates} depicts the true flow rates of the gas and oil influx in Z1 and Z2, respectively. After a spin-up period, the well flow model starts the simulation from time 10000s until time 16000s. During this period the gas flow rate undergoes two abrupt changes: one occurs at time 10600s, when the flow rate increases from 2 kg/s to 3 kg/s; the other takes place at time 12800s, when the flow rate reduces from 3 kg/s to 1 kg/s. The oil flow rate also experiences one jump from 10 kg/s to 15 kg/s, which happens at time 10600s. In accordance with the above settings, Fig. \ref{subfig: temp_press} also shows the pressure and temperature recorded by the gauges placed near Z1 and Z2, respectively (see Fig. \ref{fig:well_example}). These data are collected every 120s, and are contaminated with the Gaussian noise $N(\mathbf{v}_k; \mathbf{0},\mathbf{W}_k)$ (cf. Eq. (\ref{eq:y_obv})). Here $\mathbf{W}_k$ is a diagonal matrix, meaning that we assume the observation errors of pressure and temperature are uncorrelated with each other. In the experiment, the standard deviations of the measurement noise are $0.02\%$ of the magnitudes of the measured pressure and temperature data.         

The configuration of the Markov jump process is the same as that in \citet{LorentzenSPEEstimation}. Specifically, for both the gas and oil flow rates, the multipliers $\theta$ in Eq. (\ref{eq:rates_model}) take values from the same finite set $\{0.5,0.75,1,1.25,1.5\}$, while the corresponding probabilities of taking those values are specified by the set  $\{0.1,0.1,0.6,0.1,0.1\}$. The APF is initialized with the true flow rates at time 10000s multiplied by some samples of $\theta$ drawn at random.  

With the above said, in what follows we present the numerical results of the APF, with the variances of the Markov jump process being manually tuned as in \citet{LorentzenSPEEstimation}, and those obtained from the fix interval/lag smoothing approaches. In all these cases, the sample size of the particles is 500.

\subsubsection{Results with the manually tuned variances}

For reference, we first report the results of the APF with manually tuned variances. Following \citet{LorentzenSPEEstimation}, we let the variances of the gas and oil flow rates be both at 0.5. These manually chosen variances are used in the Markov jump process Eq. (\ref{eq:rates_model}), based on which the flow rate samples (particles) are drawn at random. The generated samples are then taken as the inputs to the well flow model, and the corresponding outputs of temperature and pressure are compared to those recorded by the gauges. Based on the data misfit between the simulated temperature and pressure and those from the gauges, the APF will adjust the relative weights of the particles, so that the weighted mean of the particles may match the observed temperature and pressure better. Since a set of samples is used, it also allows one to evaluate the uncertainties in the process of estimation.    

\begin{figure*}
\centering
\subfigure[Flow rates]{ \label{subfig:fig_par_fit_no_EM_POS}
\includegraphics[scale=0.4]{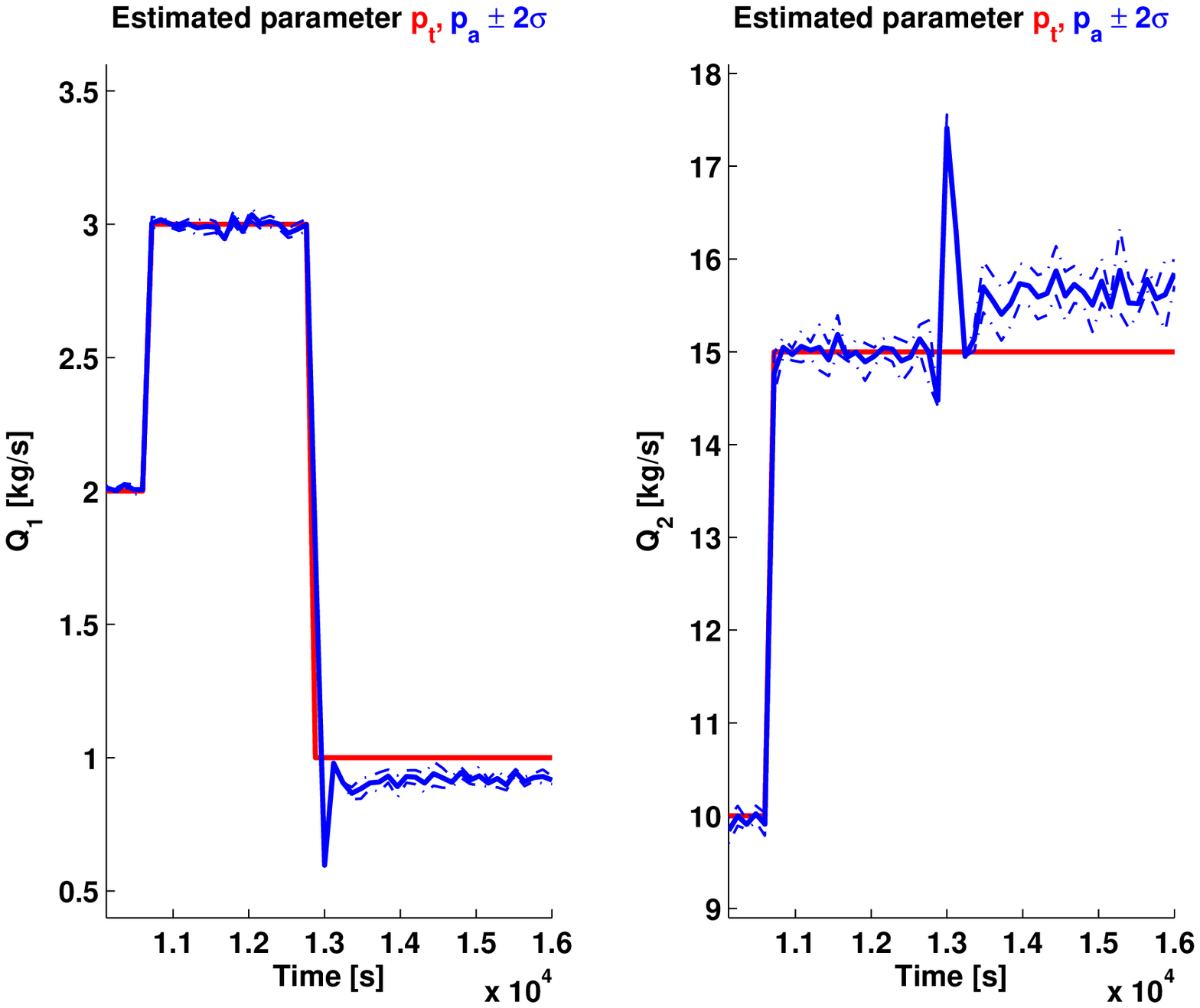}
}
\subfigure[Pressure and temperature]{ \label{subfig:fig_obs_fit_no_EM_POS}
\includegraphics[scale=0.4]{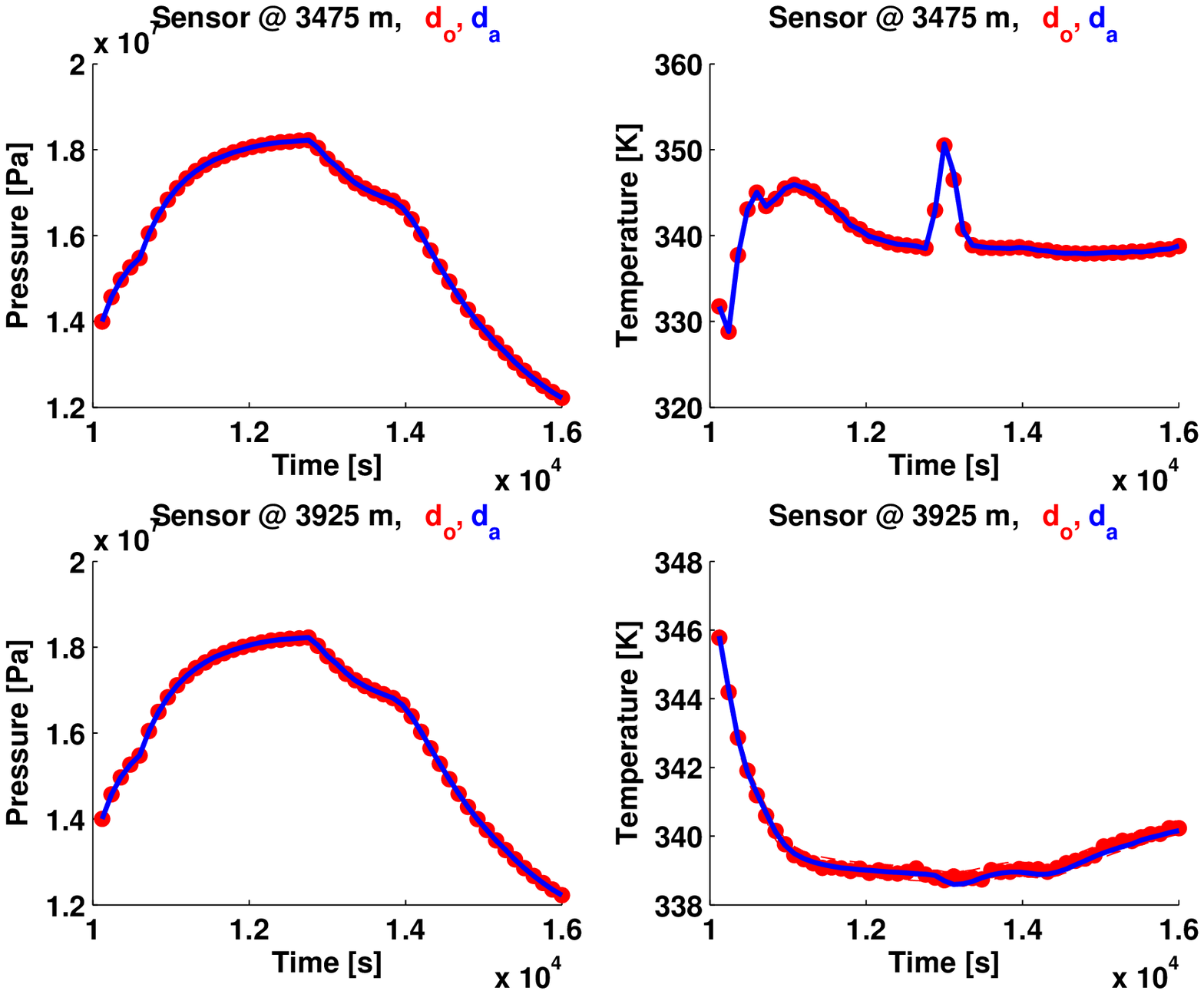}
}
\caption{\label{fig:performance_EM_POS} Performance of the APF with manually tuned variances. (a) Profiles of the gas and oil flow rates in Z1 and Z2. The true ones are in red, and the estimated ones are in blue (solid). Also plotted in blue (dash-dotted) are the flow rates that are $\pm$2 standard deviations away from the estimated ones; (b) Pressure and temperature. Those recorded by the gauges are in red (dotted), while those simulated ones based on the estimated flow rates are in blue (solid).}
\end{figure*} 

Fig. \ref{fig:performance_EM_POS} shows the estimated flow rates (panel (a)) and the corresponding simulated temperature and pressure (panel (b)). As one can see in Fig. \ref{subfig:obs_fitting_FInterval}, the simulated measurements match those recorded by the gauges well. On the other hand, before the second abrupt change of the gas flow rate happens (at time 12800s), the estimated gas and oil flow rates also match the true ones well. However, after the abrupt change at time 12800s, the estimated flow rates appear to deviate from the true ones. This possibly reflects the relative cumbersomeness, to some extent, of using constant variances in the situations with rapid changes in the fluid dynamics. The deviation of the estimated flow rates from the true ones, in terms of the root mean squared errors (RMSEs) averaged over the time interval between 10000s and 16000s, is 0.2975. 

\subsubsection{Results with the fixed interval smoothing approach} 

\begin{figure*}
\centering
\subfigure[Flow rates]{ \label{subfig:par_fitting_FInterval}
\includegraphics[scale=0.4]{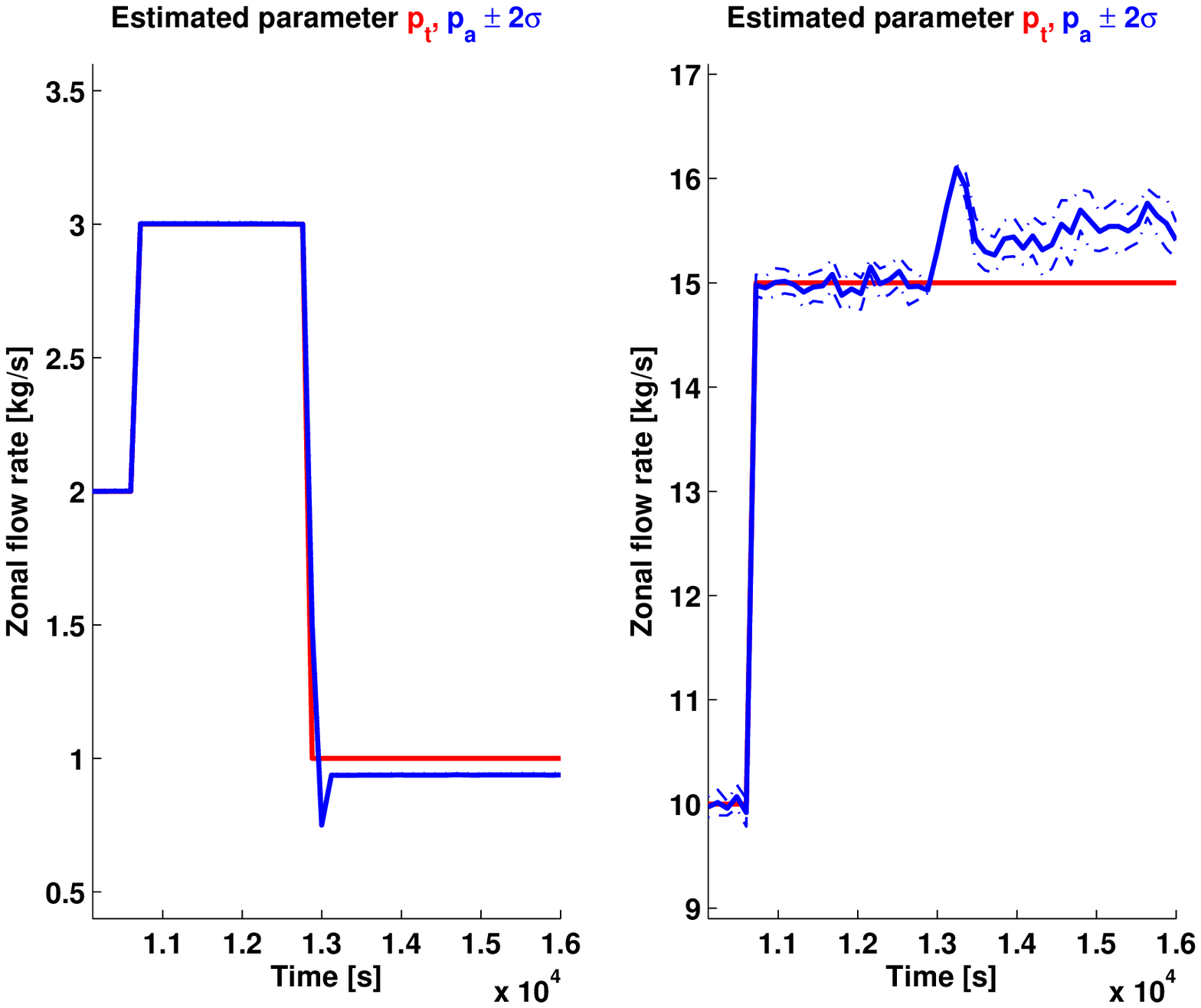}
}
\subfigure[Pressure and temperature]{ \label{subfig:obs_fitting_FInterval}
\includegraphics[scale=0.4]{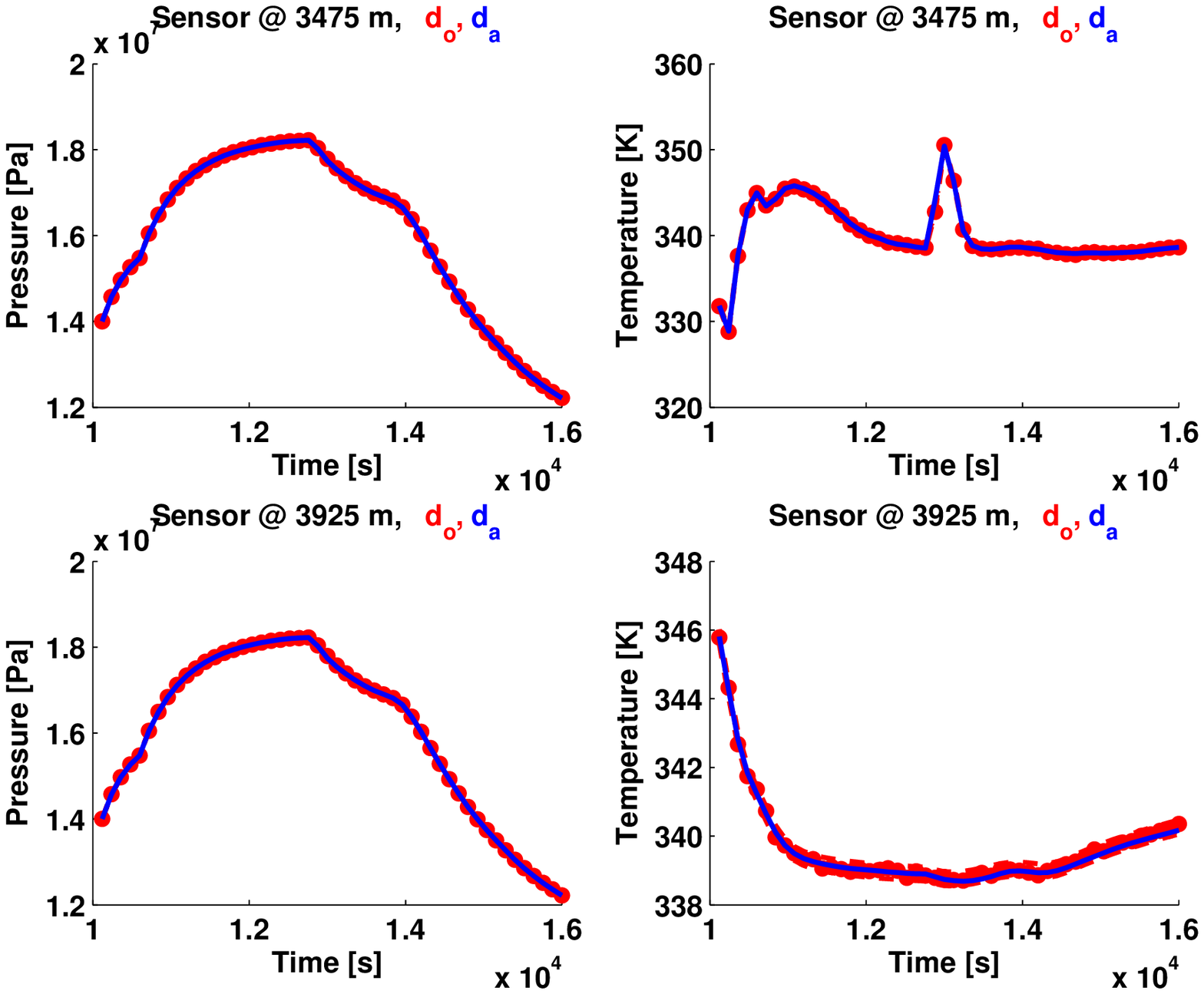}
}
\caption{\label{fig:performance_FI} 
As in Fig. \ref{fig:performance_EM_POS}, but the APF is used in conjunction with the flow rate variances estimated by the fixed interval smoothing approach. Note that in this case the spread of the gas flow rate estimates is very small and is nearly invisible in the left figure of panel (a).}
\end{figure*}

To minimize the approximate cost function in Eq. (\ref{eq:approx_cost_FI}), we adopt the MultiStart algorithm in the Global Optimization Toolbox of MATLAB$^\copyright$ R2012a, with 3 start points. The optimized variance for the gas flow rate is 0.004, while that for the oil flow rate is 0.1378. Fig. \ref{fig:performance_FI} shows the estimated flow rates (panel (a)) and the corresponding simulated temperature and pressure (panel (b)), when the APF is used in conjunction with the flow rate variances optimized through the fixed interval smoothing approach. With the constant estimated variances, the assimilation results in Fig. \ref{fig:performance_FI} appear similar to those in Fig. \ref{fig:performance_EM_POS}. However, its corresponding time mean RMSE becomes 0.2216, lower than that in Fig. \ref{fig:performance_EM_POS}.         

\subsubsection{Results with the lag-1 smoothing approach} 

\begin{figure*}
\centering
\includegraphics[scale=0.5]{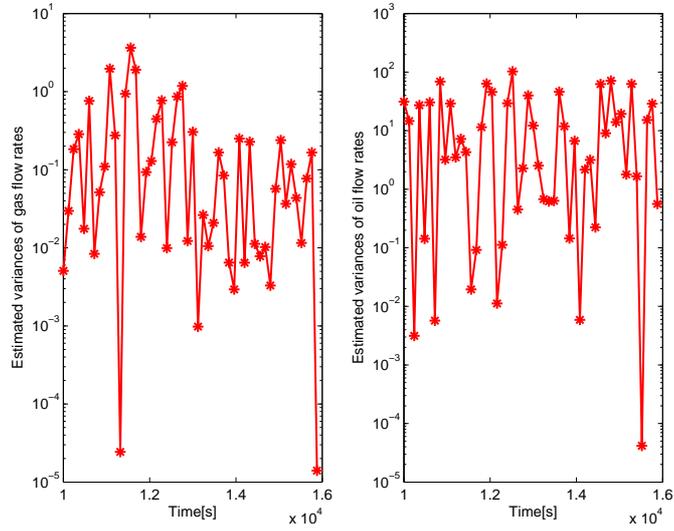}

\caption{\label{fig:estimated_variances_EM} Time series of the estimated variances of the gas (left) and oil (right) flow rates.}
\end{figure*}

Now we examine the performance of the APF when it is used in conjunction with the lag-1 smoothing approach. To implement the iteration process of the EM algorithm, the initial variances of both the gas and oil flow rates are set to 1. The stopping condition is either (a) the relative Euclidean norm change is less than $10^{-3}$, or (b) the iteration number reaches 100. In the experiments, it is found that the iteration process typically stops after 3--4 iterations, meaning that the variances estimate converges quite fast. For illustration, in Fig. \ref{fig:estimated_variances_EM} we report the (final) estimated variances of the gas (left) and oil (right) flow rates as functions of time. As one can see there, the estimated variances oscillate with time frequently, and may cross different orders of magnitudes in a short time. One consequence of such oscillations is that the corresponding estimated flow rates may appear less smooth in comparison with the fixed interval smoothing approach, as will be shown later. 

\begin{figure*}
\centering
\subfigure[Flow rates]{ \label{subfig:par_fitting_FLag}
\includegraphics[scale=0.4]{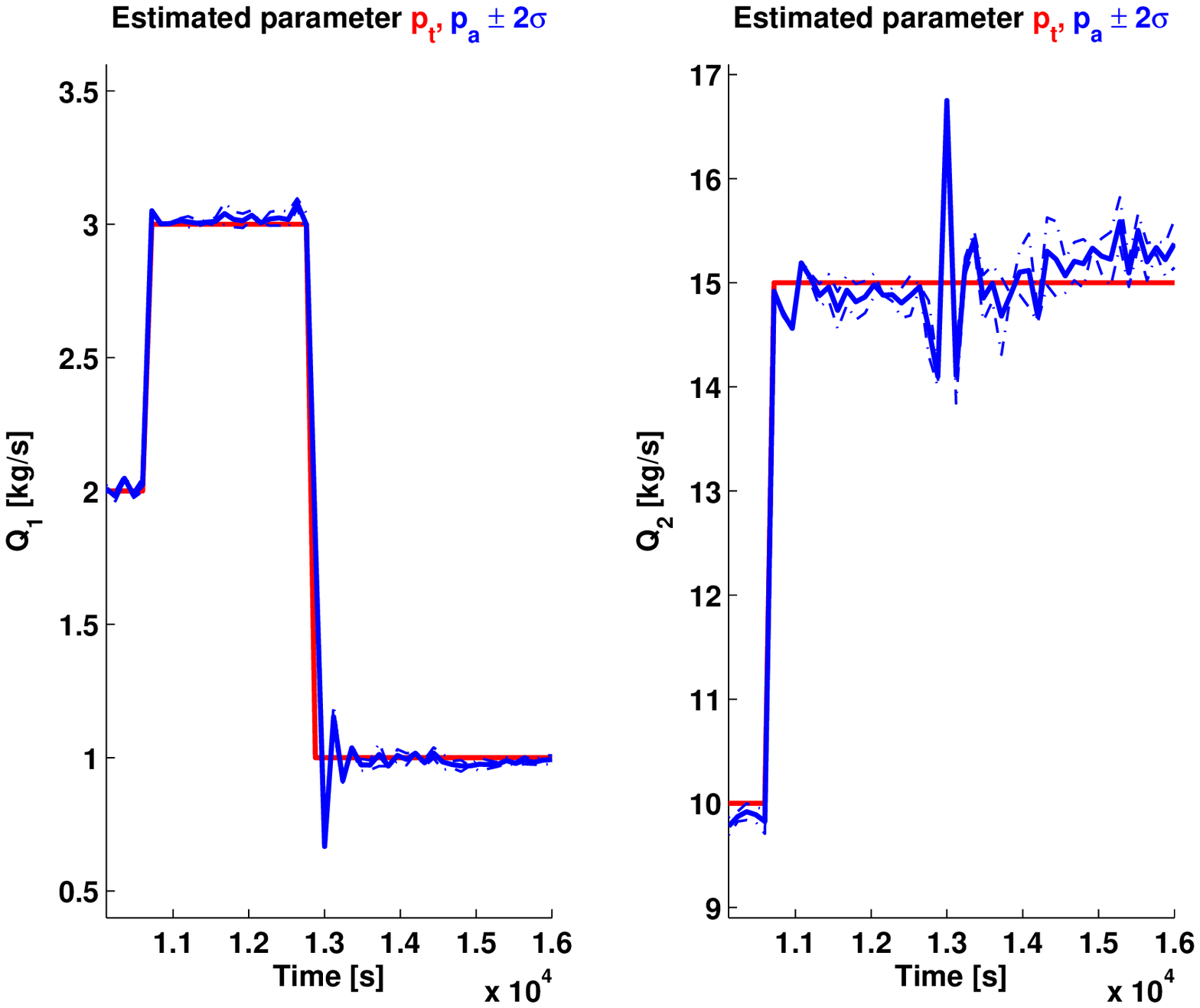}
}
\subfigure[Pressure and temperature]{ \label{subfig:obs_fitting_FLag}
\includegraphics[scale=0.4]{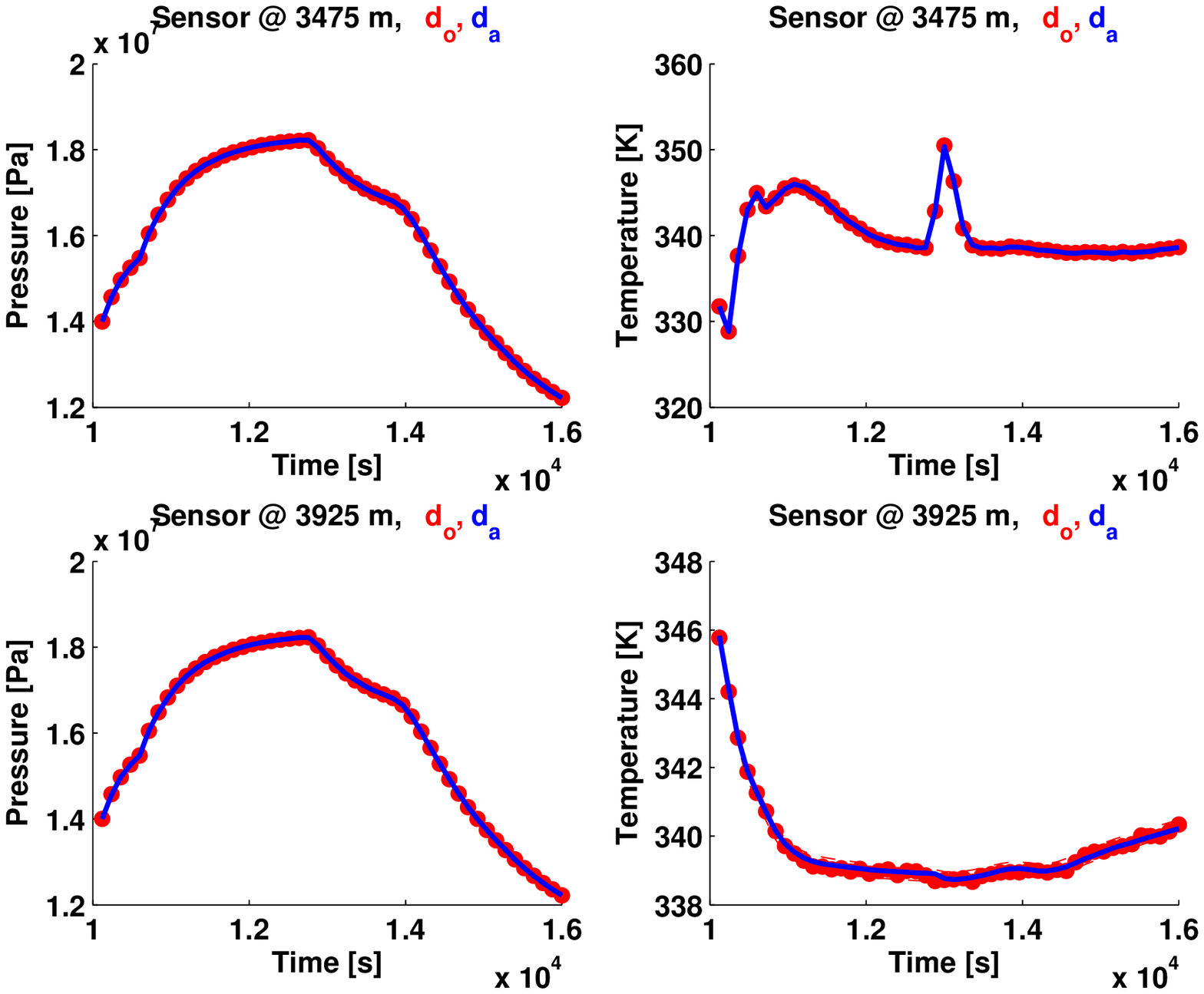}
}
\caption{\label{fig:performance_FLag} As in Fig. \ref{fig:performance_EM_POS}, but now the APF is used in conjunction with the flow rate variances estimated by the lag-1 smoothing approach.}
\end{figure*}  
    
For the APF equipped with the lag-1 smoothing approach, its estimated flow rates and the corresponding simulated temperature and pressure are reported in Figs. \ref{subfig:par_fitting_FLag} and \ref{subfig:obs_fitting_FLag}, respectively. Comparing Figs. \ref{subfig:obs_fitting_FInterval} and \ref{subfig:obs_fitting_FLag}, one can see the APF with both smoothing approaches yield close performance in terms of matching the temperature and pressure of the gauges. On the other hand, the APF with the lag-1 smoothing approach exhibits different behavior from that with the fixed interval smoothing approach, in terms of the estimated flow rates. Because the variances in the lag-1 smoothing approach oscillate with time, the estimated flow rates tend to appear less smooth than those in the fixed interval smoothing approach. For the same reason, though, after the abrupt change at time 12800s, the APF with the lag-1 smoothing approach exhibits better ability to track the true flow rates, as can be seen in Fig. \ref{subfig:par_fitting_FLag}. Overall, when the APF is equipped with the lag-1 smoothing approach, the corresponding time mean RMSE becomes 0.1916, further lower than that in the fixed interval approach. 

\subsection{The imperfect observation scenario (IOS)}

\begin{figure*}
\centering
\includegraphics[scale=0.5]{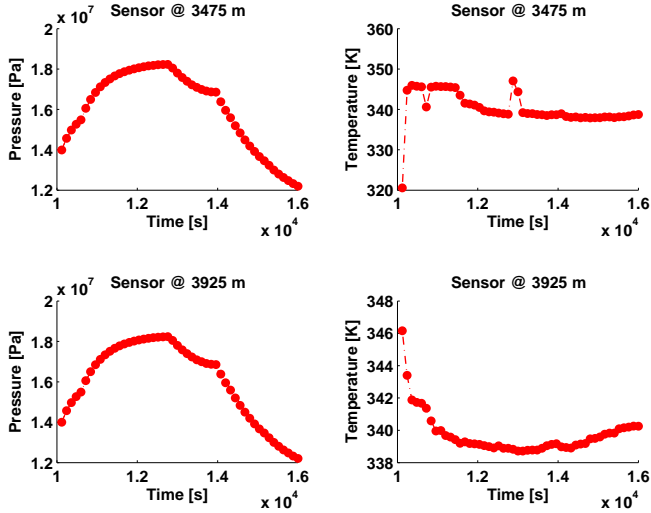}
\caption{\label{synthetic_obs_IOS} Measured pressure and temperature from the gauges, generated by the well flow model with the finer resolution (5m).}
\end{figure*} 

In a recent work \citep{Lorentzen2014estimation}, the framework proposed in \citet{LorentzenSPEEstimation} is applied to study two full-scale field cases, in which various uncertainties (e.g., observation errors) in data assimilation are involved. For our purpose here, we consider a simplified IOS in which certain imperfection may be introduced to the synthetic observation system (i.e., the well flow model) in order to evaluate the performance of the APF with the variances obtained through different ways. The experimental settings in this section are the same as those in the POS, except for the potential imperfection in the observation system resulting from the following two sources of uncertainties. One is that the well flow model used to generate the synthetic observations has a resolution different from that of the well flow model adopted in data assimilation. Specifically, to generate the synthetic observations, the spatial resolution of the well flow model is 5m (in md) along the well trajectory. While in data assimilation, the spatial resolution of the well flow model is 50m instead. For comparison, in Fig. \ref{synthetic_obs_IOS} we show the measurements generated by the well flow model with the finer resolution (the true gas and oil flow rates are the same as those in Fig. \ref{subfig: flowrates}), which are used in all our experiments below. Comparing Figs. \ref{subfig: temp_press} and \ref{synthetic_obs_IOS}, there are some clear differences in the temperature data.  

The other source of uncertainties is that the pdf of the observation error $\mathbf{v}_k$ in Eq. (\ref{eq:y_obv}) may be mis-specified. Concretely, let $\mathbf{W}_k$ be the diagonal covariance matrix that is used to generate the synthetic observations with the finer resolution (5m) in the well flow model (here $\mathbf{W}_k$ is constructed in the same way as in the POS). In data assimilation we consider three cases, with the observation error covariance matrix being $0.5 \times \mathbf{W}_k$, $1 \times \mathbf{W}_k$ and $2 \times \mathbf{W}_k$, respectively. Therefore in the cases with $0.5 \times \mathbf{W}_k$ and $2 \times \mathbf{W}_k$, the observation error covariance matrices are mis-specified. 

Fig. \ref{fig:performance_IOS} shows the performance of the APF when the observation error covariance matrix is $1 \times \mathbf{W}_k$. The upper panels report the estimated flow rates (left) and the corresponding simulated observations (right), respectively, when the APF uses the manually chosen variances (0.5 for both gas and flow rates) in \citet{LorentzenSPEEstimation}. Similarly, the middle and lower panels present the results of the APF when the fixed-lag and lag-1 smoothing approaches are adopted to optimize the variances. As one can see in Fig. \ref{fig:performance_IOS}, the APF can still match the pressure data well in all cases. However, its matching of the temperature data is worsened compare to the cases in the POS. Accordingly, the corresponding estimated flow rates also exhibit relatively large estimation errors. Indeed, with the manually chosen variances, the time mean RMSE becomes 1.5228, while those with the variances optimized by the fixed-lag and lag-1 smoothing approaches are 0.9321 and 1.3533, respectively. Therefore, in terms of estimation accuracies, the APF with manually chosen variances under-performs those with the variances optimized by both smoothing approaches, and in this particular case the fixed-lag smoothing approach leads to the lowest time mean RMSE.    
        
\newcommand{\nScale}{0.3}
\begin{figure*}
\centering
\subfigure[Flow rates (manual)]{ \label{subfig:fig_par_fit_noEM_IOS_1xW}
\includegraphics[scale=\nScale]{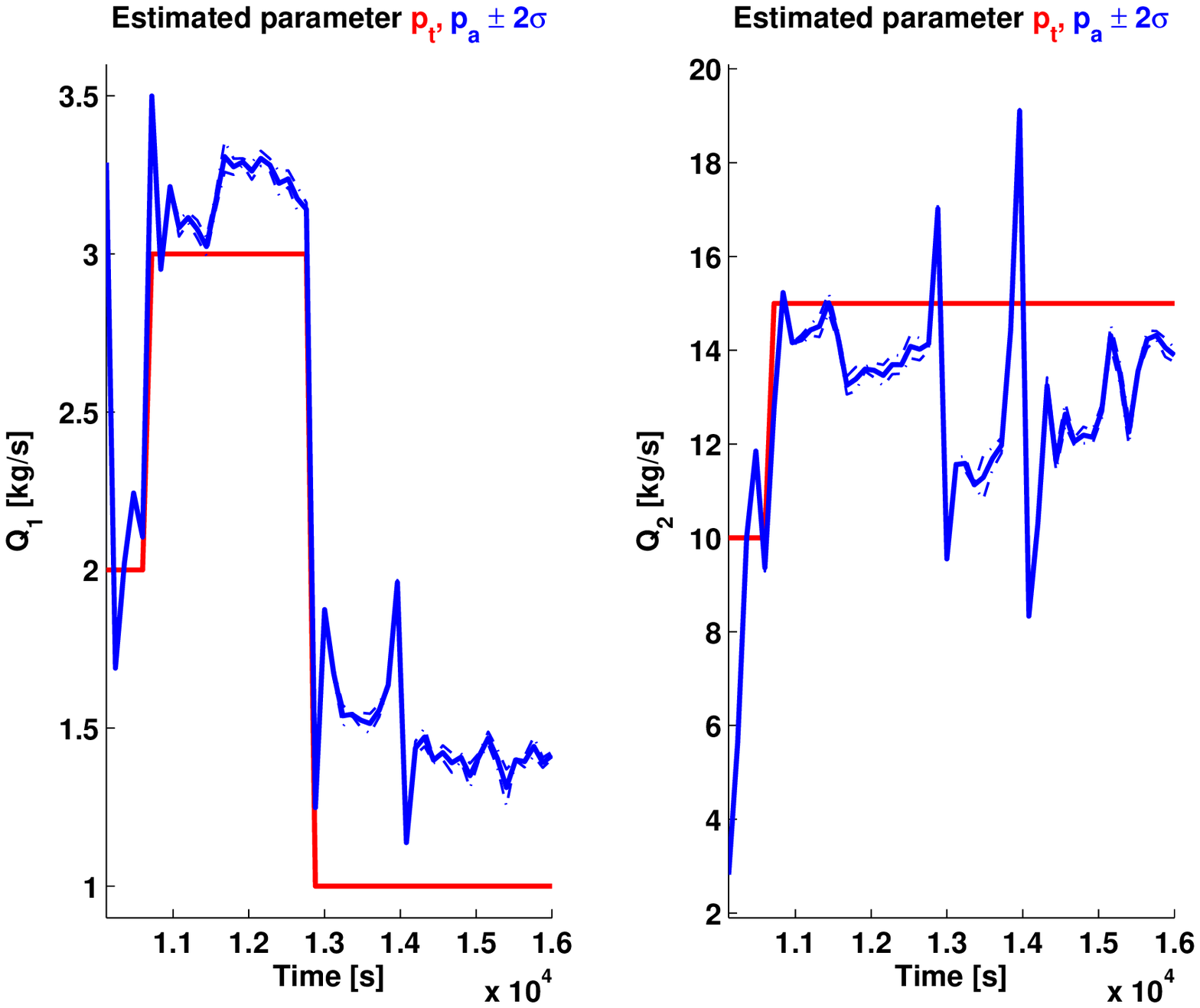}
}
\subfigure[Pressure and temperature (manual)]{ \label{subfig:fig_obs_fit_noEM_IOS_1xW}
\includegraphics[scale=\nScale]{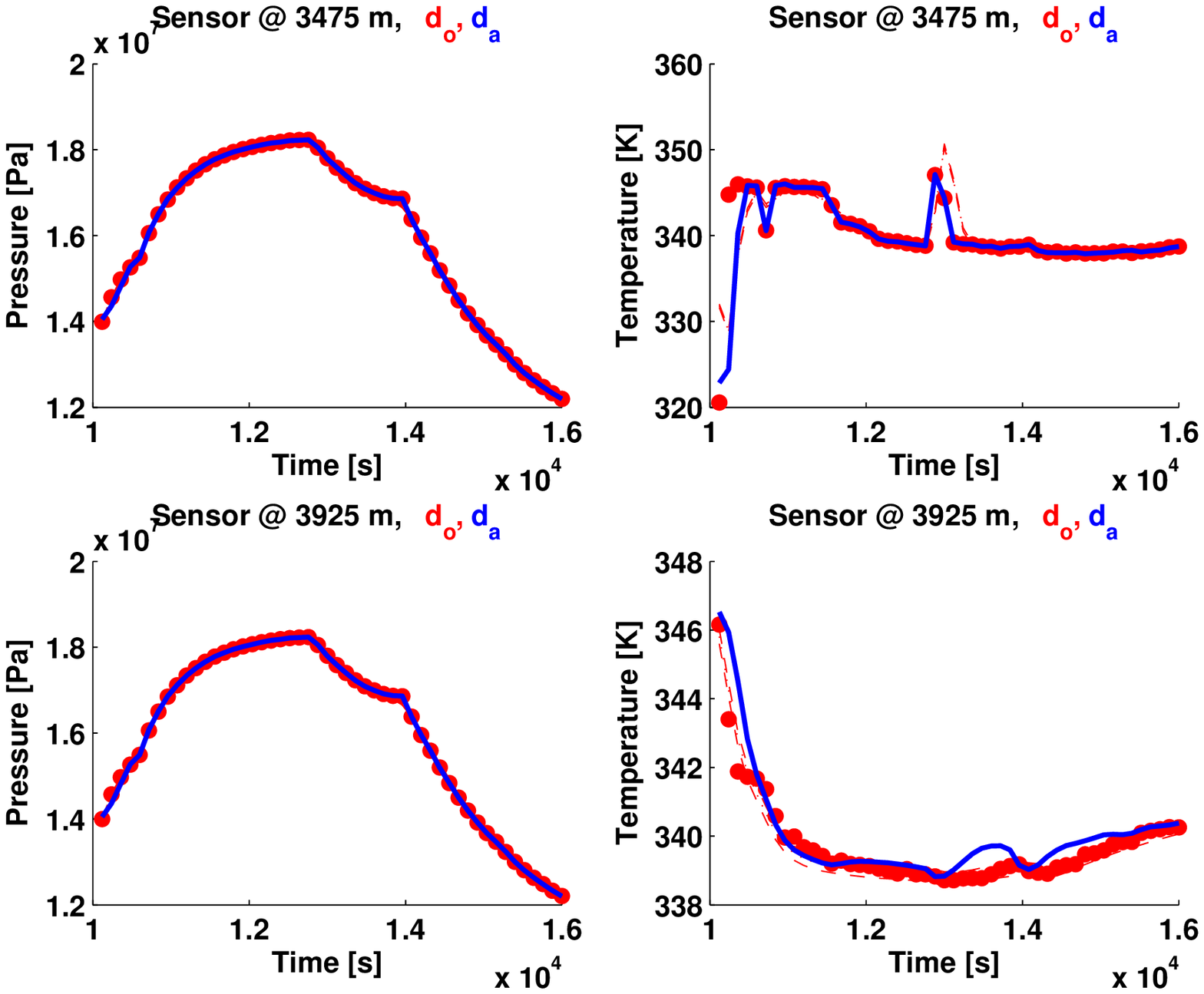}
}

\subfigure[Flow rates (lag-1)]{ \label{subfig:fig_par_fit_globalOpt_IOS_1xW}
\includegraphics[scale=\nScale]{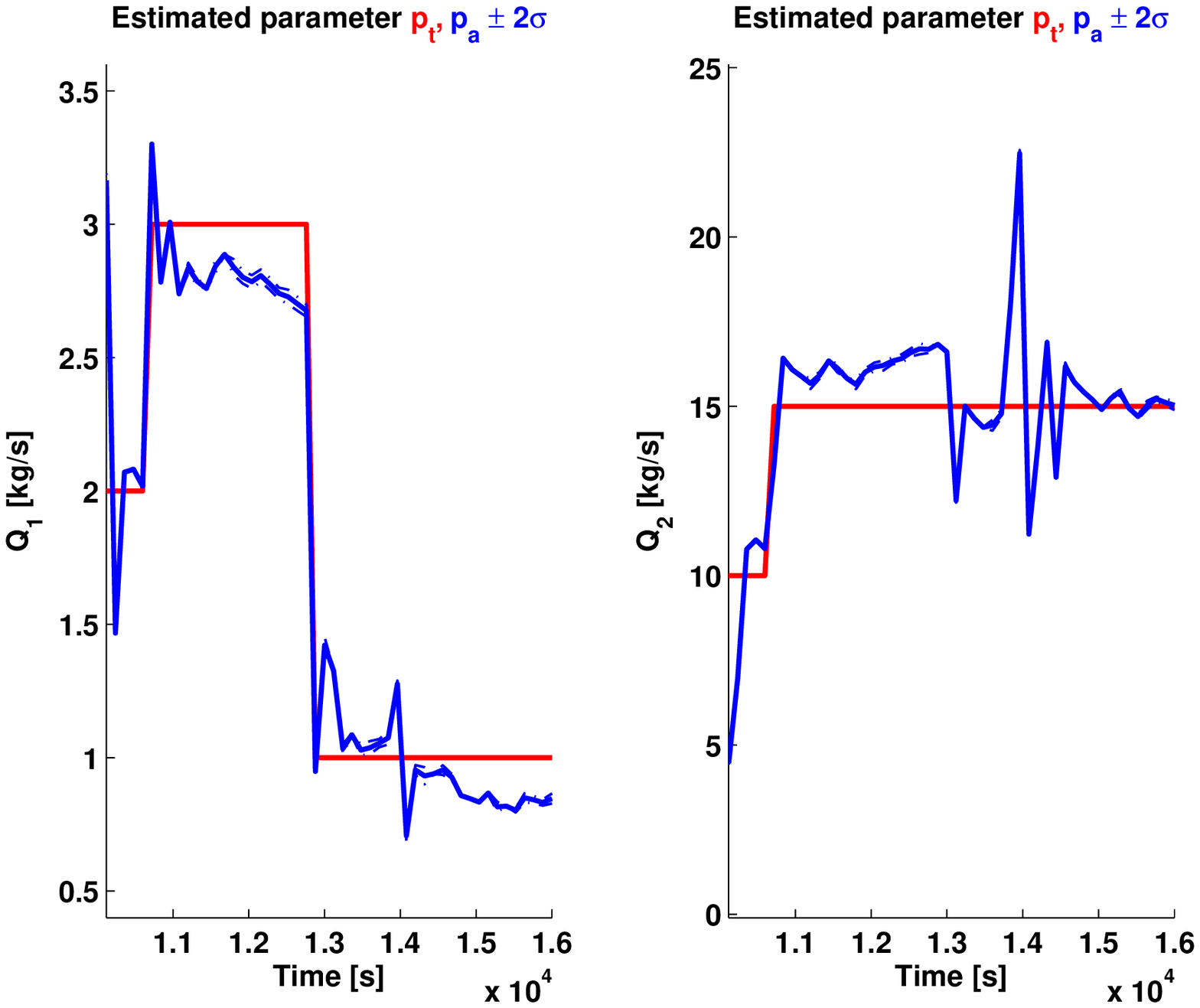}
}
\subfigure[Pressure and temperature (lag-1)]{ \label{subfig:fig_obs_fit_globalOpt_IOS_1xW}
\includegraphics[scale=\nScale]{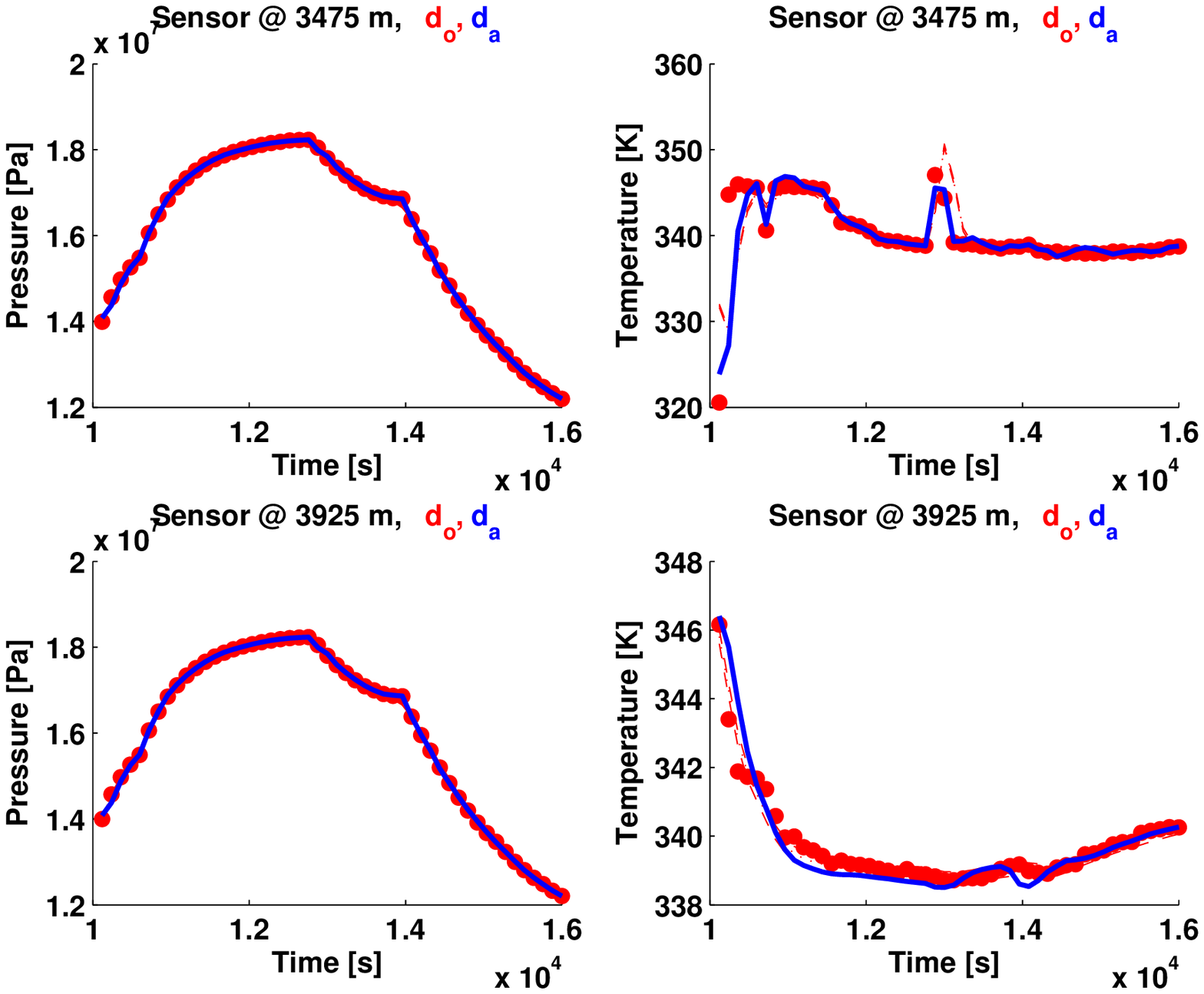}
}

\subfigure[Flow rates (fixed-lag)]{ \label{subfig:fig_par_fit_EM_IOS_1xW}
\includegraphics[scale=\nScale]{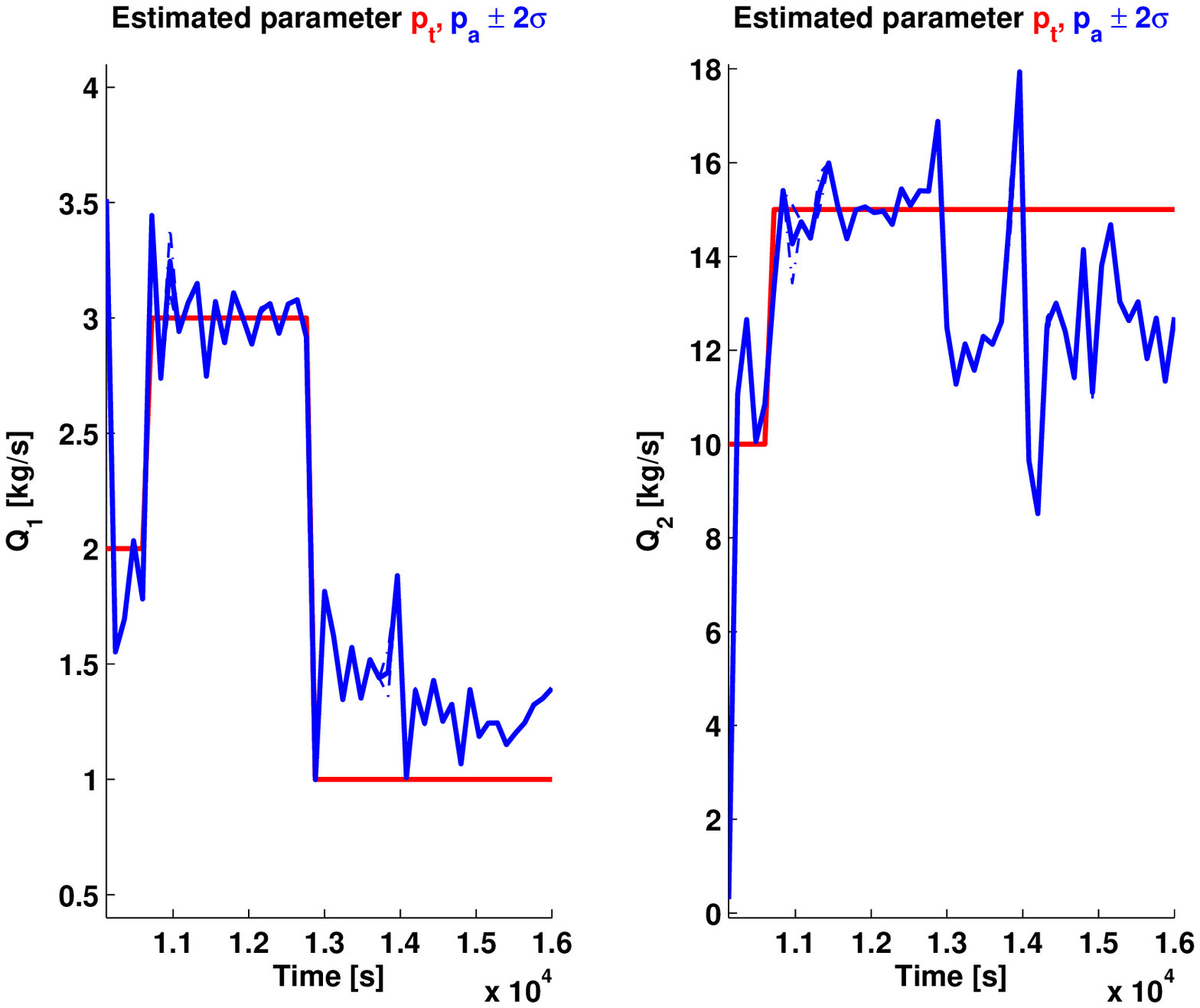}
}
\subfigure[Pressure and temperature (fixed-lag)]{ \label{subfig:fig_obs_fit_EM_IOS_1xW}
\includegraphics[scale=\nScale]{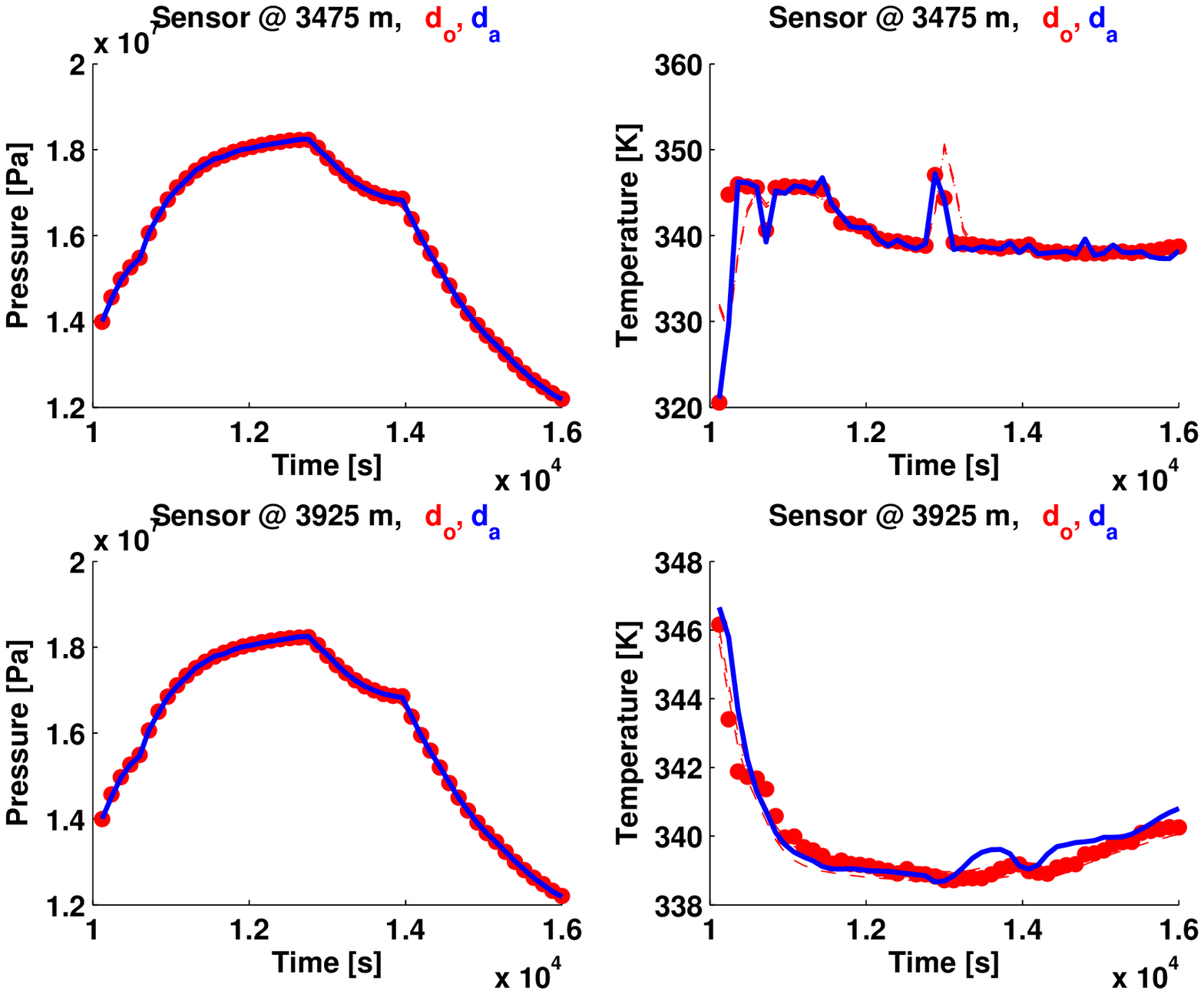}
}
\caption{\label{fig:performance_IOS} Upper panels (a,b): Performance of the APF with the manually chosen variances (0.5 for both the oil and gas flow rates). Middle panels (c,d): Performance of the APF with the variances (0.1270 for the gas flow rate and 0.1075 for the oil flow rate) optimized by the fixed-lag smoothing approach. Lower panels (e,f): Performance of the APF with the variances optimized by the lag-1 smoothing approach.}
\end{figure*} 

For brevity, the experiment results in terms of time mean RMSEs, with the observation error covariance matrix being $0.5 \times \mathbf{W}_k$ and $2 \times \mathbf{W}_k$, respectively, are summarized in Table \ref{table:performance_IOS}. As one can see, in both cases, the APF with manually chosen variances again under-performs the one with the variances optimized by the two smoothing approaches. In the case with $0.5 \times \mathbf{W}_k$, the lag-1 smoothing approach seems to result in the best estimation, while in the case with $2 \times \mathbf{W}_k$, it is the fixed-lag smoothing approach that appears to have the lowest time mean RMSE.      

Compared to the results in the POS (cf. Figs. \ref{fig:performance_EM_POS} -- \ref{fig:performance_FLag}), the performance of the APF deteriorates in the IOS for all three observation error covariance matrices used in data assimilation. The reason of the performance downgrade may be largely attributed to the imperfection in the observation system, since in the Bayesian estimation framework, it is essential for one to know the statistical distributions of both model and observation errors (see, for example, Eqs. (\ref{BRR:prediction}) and (\ref{BRR:update})), while the imperfection in the observation system may make the estimation become sub-optimal. In such circumstances, it may be desirable for one to adopt robust data assimilation algorithms (see, for example, \citealp{Luo2014-efficient,Luo2011_EnLHF,Simon2006}) that are less sensitive to the precise knowledge of the model and/or observation error distributions. An investigation in this aspect is, however, beyond the scope of the current work.  

\begin{table}[!t]
\caption{ \label{table:performance_IOS} Time mean RMSE of the APF when the observation error covariance matrix in assimilation is $0.5 \times \mathbf{W}_k$, $1 \times \mathbf{W}_k$ and $2 \times \mathbf{W}_k$, respectively.} 
\begin{center}
\vspace{-0.5cm}
\begin{tabular}{lccc}
\hline
\hline
 & manual & fixed interval & lag-1 \\
 \hline
 Time mean RMSE ($0.5 \times \mathbf{W}_k$) & 1.5026 & 1.4869 & 1.1885 \\
 Time mean RMSE ($1 \times \mathbf{W}_k$) & 1.5228 & 0.9321 & 1.3533 \\
 Time mean RMSE ($2 \times \mathbf{W}_k$) & 1.8201 & 1.2431 & 1.5020 \\
 \hline
 \hline
\end{tabular}
\end{center}
\end{table}  


\section{Conclusion} \label{sec:conclusion}

In this work we studied the soft-sensing problem based on a previously established framework. The salient features of this framework include the transient, instead of steady-state, well flow model, a Markov jump process that has better ability in capturing abrupt changes in the fluid dynamics, and a modern estimation approach, the auxiliary particle filter (APF), that guides the estimated flow rates toward matching the temperature and pressure from the physical sensors.

This work focused on estimating the variances of the flow rates in the Markov jump process. To this end, we proposed and implemented two approaches. One utilizes all the available measurements, e.g., temperature and pressure, in a fixed time interval, and is thus called the fixed interval smoothing approach. The other uses the measurements up to (and including) 1-step ahead of the time instant at which the variances of the flow rates are to be estimated, and therefore is named the lag-1 smoothing approach. In our implementations, the fixed interval smoothing approach provides constant variance estimates to the APF, and may thus tend to produce more smooth flow rate estimation than the lag-1 smoothing approach. On the other hand, the APF with the lag-1 smoothing approach requires less waiting time for the necessary measurements to arrive, and tends to run faster in terms of computational time.     

In a numerical example, we investigated the performance of our framework in both the perfect observation scenario (POS) and the imperfect observation scenario (IOS), in which the APF ran with the manually chosen variances, and those supplied by the fixed interval and lag-1 smoothing approaches. In the POS, we showed that when the APF is used in conjunction with the variances from either of the three methods, the simulated temperature and pressure of the estimated flow rates exhibit good matching to those of the physical sensors. In terms of estimation accuracies of the flow rates, however, the APF with the variances optimized by either smoothing approach seems to outperform the one with manually chosen variances. Similar results were also observed in the IOS. However, due to the imperfection in the observation system, the performance of the APF tends to deteriorate, in terms of both the estimation accuracy and the ability to match the observations. These results thus suggest the importance for one to have an observation system that is able to model the underlying physics reasonably well.  
    
\section*{Acknowledgment} 
The authors would like to thank two anonymous reviewers for their constructive comments and suggestions that have significantly improved the work. We also acknowledge financial supports from the Research Council of Norway, ConocoPhillips Skandinavia AS and GDF SUEZ E\&P Norge AS, through the project ``Transient well flow modelling and modern estimation techniques for accurate production allocation.''

\appendix
\numberwithin{equation}{section}
\section{The auxiliary particle filter} \label{sec:appendix}
Here we provide an outline of the main procedures in the auxiliary particle filter (APF), following \citet{Luo2014-efficient,Pitt_ASIR}. For illustration, consider the state estimation problem in the following system
\begin{linenomath*}
\begin{subequations} \label{ps}
\begin{align}
 \label{ps_dyanmical_system} & \mathbf{Q}_k  = \mathcal{M}_{k,k-1} \left( \mathbf{Q}_{k-1}\right) + \mathbf{u}_k \, ,  \\
  \label{ps_observation_system} &  \mathbf{y}_k  = \mathcal{H}_{k} \left( \mathbf{Q}_{k}\right) + \mathbf{v}_k \, .
\end{align}
\end{subequations}
\end{linenomath*}
Here, $\mathbf{Q}_{k} \in \mathbb{R}^{r}$ is the $r$-dimensional model state at time instant $k$, $\mathbf{y}_{k} \in \mathbb{R}^{p}$ the corresponding observation of $\mathbf{Q}_{k}$, $\mathbf{u}_k \in \mathbb{R}^{r}$ the model error with the associated pdf $p \left( \mathbf{u}_k \right)$, and $\mathbf{v}_k \in \mathbb{R}^{p}$ the observation error with the associated pdf $p \left( \mathbf{v}_k \right)$. The transition operator $\mathcal{M}_{k,k-1}: \mathbb{R}^{r} \rightarrow \mathbb{R}^{r}$ maps $\mathbf{Q}_{k-1}$ to $\mathbf{Q}_{k}$ (and it can be a stochastic map as in Eq. (\ref{eq:rates_model})), and the observation operator $\mathcal{H}_{k}: \mathbb{R}^{r} \rightarrow \mathbb{R}^{p}$ projects $\mathbf{Q}_{k}$ from the state space onto the observation space. The problem of our interest is to estimate the posterior pdf of the model state $\mathbf{Q}_k$ at time instant $k$, given the observations $\mathbf{Y}_{k} = \left \{ \mathbf{y}_{k},  \mathbf{y}_{k-1}, \dotsb \right \} $ up to and including $k$, together with the prior pdf $p \left( \mathbf{Q}_{i} | \mathbf{Y}_{i-1} \right)$ of the model state $\mathbf{Q}_i$ at some earlier instant $i$ ($i \le k$). 

Recursive Bayesian estimation (RBE) \citep{Arulampalam2002} provides a probabilistic framework that recursively solves the state estimation problem in terms of some conditional pdfs. Let $p \left( \mathbf{Q}_{k} | \mathbf{Y}_{k-1} \right)$ be the prior pdf of $\mathbf{Q}_{k}$ conditioned on the observations $\mathbf{Y}_{k-1}$ up to and including time $k-1$, but without the knowledge of the observation $\mathbf{y}_{k}$ yet. Once the observation $\mathbf{y}_k$ is known, one incorporates the information content of $\mathbf{y}_k$ according to Bayes' rule to update the prior pdf to the posterior one $p \left( \mathbf{Q}_{k} | \mathbf{Y}_{k} \right)$. By evolving $p \left( \mathbf{Q}_{k} | \mathbf{Y}_{k} \right)$ forward in time, one obtains a prior pdf $p \left( \mathbf{Q}_{k+1} | \mathbf{Y}_{k} \right)$ at the next time instant. Concretely, the mathematical description of RBE consists of \citep{Arulampalam2002}:\\
\noindent Prediction step:
\begin{linenomath*}
\begin{equation} \label{BRR:prediction}
p \left( \mathbf{Q}_{k} | \mathbf{Y}_{k-1} \right) =  \int p \left( \mathbf{Q}_{k} | \mathbf{Q}_{k-1} \right)  p \left( \mathbf{Q}_{k-1} | \mathbf{Y}_{k-1} \right) d\mathbf{Q}_{k-1} \, ,
\end{equation} \\
\noindent and filtering step:
\begin{linenomath*}
\begin{equation} \label{BRR:update}
p \left( \mathbf{Q}_{k} | \mathbf{Y}_{k} \right) =  \dfrac{ p \left( \mathbf{y}_{k} | \mathbf{Q}_{k}  \right) p \left( \mathbf{Q}_{k} | \mathbf{Y}_{k-1} \right) }{\int p \left( \mathbf{y}_{k} | \mathbf{Q}_{k}  \right) p \left( \mathbf{Q}_{k} | \mathbf{Y}_{k-1} \right)  d\mathbf{Q}_{k}} \, ,
\end{equation}
\end{linenomath*}
\end{linenomath*}
where the transition pdf $p \left( \mathbf{Q}_{k} | \mathbf{Q}_{k-1}  \right)$ and the likelihood function $p \left( \mathbf{y}_{k} | \mathbf{Q}_{k}  \right)$ are assumed known, in light of the knowledge of the pdfs of the model and observation errors in Eq.~(\ref{ps}). Once the explicit forms of the conditional pdfs in  Eqs.~(\ref{BRR:prediction}) and (\ref{BRR:update}) are obtained, the optimal estimate and other associated statistical information can be derived based on a certain optimality criterion, e.g., minimum variance or maximum likelihood. Thus RBE provides a solution of the estimation problem, and conceptually leads to the optimal nonlinear filter.

In practice, however, difficulties often arise in deriving the exact optimal filter, largely due to the fact that the integrals in Eqs.~(\ref{BRR:prediction}) and (\ref{BRR:update}) are often intractable. Therefore one may have to adopt a certain approximation scheme for evaluation. In many situations, it if often reasonable to approximate the prior and posterior pdfs by certain Gaussian distributions. With this approximation, the RBE framework reduces to the KF or certain extensions in nonlinear systems, e.g., the extended Kalman filter (EKF, see \citealp{Jazwinski1970,Simon2006}), the sigma point Kalman filters (SPKF, see, for example, \citealp{Ito-gaussian,Julier2000,Luo-ensemble,Luo2012-ddf,Norgaard--new}), and the ensemble Kalman filter (EnKF, see, for example, \citealp{Anderson-ensemble,Bishop-adaptive,Burgers-analysis,Luo2012-residual,luo2013-covariance,Pham2001,Whitaker-ensemble}). 
In addition, a family of nonlinear and non-Gaussian data assimilation algorithms based on the Gaussian mixture model (GMM) approximation, often called the Gaussian sum filter (see, for example, \citealp{Alspach-nonlinear,Chen2000-mixture,Hoteit2012,Hoteit2008,Luo2008-spgsf1,Sorenson-recursive,Stordal-bridging-2010}), can be constructed as a bank of parallel nonlinear Kalman filters (e.g., the EKF, the SPKF, or the EnFK), which enhances the re-usability of existing codes of nonlinear Kalman filters. 

Specific to our focus in the current work, if the Monte Carlo approximation is adopted to approximate the prior and posterior pdfs in the RBE framework, then it leads to the particle filter (PF, see, for example, \citealp{Doucet2001-sequential,Gordon1993,Luo2014-efficient,Pitt_ASIR}) that is also applicable to nonlinear and non-Gaussian data assimilation problems. For instance, the posterior $p \left( \mathbf{Q}_{k-1} | \mathbf{Y}_{k-1} \right)$ at the $(k-1)$th step can be approximated by
\begin{linenomath*}
\[
p \left( \mathbf{Q}_{k-1} | \mathbf{Y}_{k-1} \right) \approx \sum\limits_{i=1}^n w_{k-1,i} \delta(\mathbf{Q}_{k-1} - \mathbf{Q}^a_{k-1,i}) \, ,
\]
\end{linenomath*}
where $\mathbf{Q}^a_{k-1,i}$ ($i = 1, 2, \dotsb, n$) are the particles at the filtering step before a re-sampling algorithm (if necessary) is applied, $w_{k-1,i}^a$ are the associated weights, and $n$ is the total number of particles. For notational convenience, let $\tilde{\mathbf{Q}}^a_{k-1,i}$ be the particles generated by a re-sampling algorithm, and $\tilde{w}_{k-1,i}^a$ the corresponding weights. Suppose that we start with some particles $\mathbf{Q}^b_{k,i}$ from the prior pdf at time instant $k$, together with their associated weights $w^b_{k,i}$. Then, in consistency with Eqs.~(\ref{BRR:prediction}) and (\ref{BRR:update}), the APF has the following steps.

\begin{itemize}

\item {Filtering step:} With an incoming observation $\mathbf{y}_{k}$, the particles remain unchanged, i.e., $\mathbf{Q}^a_{k,i} = \mathbf{Q}^b_{k,i}$, while the associated weights are updated according to Bayes' rule so that
\begin{linenomath*}
\begin{equation} \label{eq:weight_update_PF}
w_{k,i}^a = \frac{w^b_{k,i} \; p(\mathbf{y}_k | \mathbf{Q}^b_{k,i})}{ \sum\limits_{i=1}^n w^b_{k,i} \; p(\mathbf{y}_k | \mathbf{Q}^b_{k,i})}  \, ,
\end{equation}
\end{linenomath*}
where $p(\mathbf{y}_k | \mathbf{Q}^b_{k,i})$ is the probability that $\mathbf{y}_k$ happens to be the observation with respect to $\mathbf{Q}^b_{k,i}$.

\item {Re-sampling step:} In high-dimensional models, weight collapse may happen and this will, in general, deteriorate the performance of the PF. To mitigate this problem, it is customary to include a re-sampling step that singles out a specified number of particles from the pool $\{\mathbf{Q}^a_{k,i} \}_{i=1}^n$ based on a certain rule. In the conventional PF, the selection is conducted based on the weights $\{w_{k,i}^a\}_{i=1}^n$ associated with $\{\mathbf{Q}^a_{k,i} \}_{i=1}^n$. However, as pointed out by \citet{Pitt_ASIR}, particles selected in this way are not guaranteed to match the future observations. As a remedy, one may take into account the particles' ability to match future observations. To this end, one can, for instance, compute the intermediate weights $\mu_{k,i} \propto w_{k,i}^a \, p(\mathbf{y}_{k+1}|\mathcal{M}_{k+1,k}( \mathbf{Q}^a_{k,i}))$, where $p(\mathbf{y}_{k+1}| \mathcal{M}_{k+1,k} ( \mathbf{Q}^a_{k,i}))$ represents the likelihood function value of the forecast $\mathcal{M}_{k+1,k} \left( \mathbf{Q}^a_{k,i}\right)$ of $\mathbf{Q}^a_{k,i}$ at time instant $k+1$. Based on $\mu_{k,i}$, one applies sequential importance re-sampling (SIR, see, for example, \citealp{Arulampalam2002}) to the indices $i=1,2,...n$ and obtains a new set of indices $i^s, s = 1,2,\dotsb,n$, to pick up samples from the set $\{\mathbf{Q}^a_{k,i}\}_{j=1}^n$. The indexed samples $\{\tilde{\mathbf{Q}}^a_{k,i^s}\}_{s=1}^n$ then has equal weights, i.e., their associated weights $\tilde{w}_{k,i^s}^a = 1/n$ for all $i^s, s = 1,2,\dotsb,n$.  

\item {Prediction step:} The re-sampled particles $\{\tilde{\mathbf{Q}}^a_{k,i^s}\}_{s=1}^n$ are propagated forward in time through the dynamical model Eq. (\ref{ps_dyanmical_system}), so as to obtain a set of particles $\{\mathbf{Q}^b_{k+1,s} \}_{s=1}^n$ at time instant $k+1$. Since at the previous re-sampling step, the likelihood function values with respect to the observation at time $k+1$ are involved in the re-sampling step, one needs to adjust the weights of $\{\mathbf{Q}^b_{k+1,s} \}_{s=1}^n$ to make them consistent with the RBE framework. To this end, the weights of $\mathbf{Q}^b_{k+1,s}$ are computed according to $w^b_{k+1,s} \propto p(\mathbf{y}_{k+1}|\mathbf{Q}^b_{k+1,s}) / p(\mathbf{y}_{k+1}|\mathcal{M}_{k+1,k} ( \tilde{\mathbf{Q}}^a_{k,i^s}))$. With these computed, one can proceed to the filtering step as described previously, and so on.
\end{itemize}

\bibliography{all_references} 
\bibliographystyle{TUPREP} 

\end{document}